\documentclass[aps,prd,twocolumn,superscriptaddress,nofootinbib]{revtex4-2}
\usepackage{amsmath,amssymb,bm}
\usepackage{graphicx}
\usepackage[colorlinks=true,linkcolor=blue!60!black,citecolor=blue!60!black,urlcolor=blue!60!black]{hyperref}
\usepackage{xcolor}
\usepackage{pgfplots}
\pgfplotsset{compat=1.16}
\usetikzlibrary{patterns}
\graphicspath{{./}}
\newcommand{\Msun}{M_\odot}
\newcommand{\xifoc}{\xi_{\rm foc}}
\newcommand{\rhodm}{\rho_{\rm dm}}
\newcommand{\vdm}{v_{\rm dm}}
\newcommand{\taurel}{\tau_{\rm rel}}
\newcommand{\MGA}{M^{\rm GA}}
\newcommand{\MBS}{M^{\rm BS}}
\newcommand{\McGA}{M_c^{\rm GA}}
\newcommand{\McBS}{M_c^{\rm BS}}
\newcommand{\tcGA}{t_c^{\rm GA}}
\newcommand{\tcBS}{t_c^{\rm BS}}
\newcommand{\trepGA}{t_{\rm rep}^{\rm GA}}
\newcommand{\trepBS}{t_{\rm rep}^{\rm BS}}
\newcommand{\RGA}{R^{\rm GA}}
\newcommand{\RBS}{R^{\rm BS}}
\newcommand{\GGA}{\Gamma^{\rm GA}}
\newcommand{\GBS}{\Gamma^{\rm BS}}
\newcommand{\rhocGA}{\rho_c^{\rm GA}}

\newcommand{\fBS}{f_{\rm BS}}
\newcommand{\Fgeo}{F_{\rm geo}}
\newcommand{\MejGA}{M_{\rm ej}^{\max,{\rm GA}}}
\newcommand{\MejBS}{M_{\rm ej}^{\max,{\rm BS}}}
\newcommand{\dGA}{\delta^{\rm GA}}
\newcommand{\dBS}{\delta^{\rm BS}}
\newcommand{\epsGA}{\epsilon^{\rm GA}}
\newcommand{\epsBS}{\epsilon^{\rm BS}}
\newcommand{\eV}{\,\mathrm{eV}}
\newcommand{\GeV}{\,\mathrm{GeV}}
\newcommand{\Gyr}{\,\mathrm{Gyr}}
\newcommand{\kpc}{\,\mathrm{kpc}}
\newcommand{\pc}{\,\mathrm{pc}}
\newcommand{\kms}{\,\mathrm{km/s}}

\begin{document}

\title{Can bosenovae deplete the galactic dark matter?}

\author{Samuel G\'omez G\'omez}
\affiliation{Departament de F\'isica, Universitat de les Illes Balears, Ctra. Valldemossa km 7.5, E-07122 Palma de Mallorca, Spain}
\author{Joshua Eby}
\affiliation{The Oskar Klein Centre, Department of Physics, Stockholm University, AlbaNova, SE-106 91 Stockholm, Sweden}

\date{\today}

\begin{abstract}
No: neither through the bosenovae of gravitational atoms formed around stars and compact objects, nor through those of boson stars formed in the galactic dark-matter halo. We assume that all the dark matter consists of ultralight bosons with attractive self-interactions, which can be captured in the gravitational potential of astrophysical objects or condense into self-gravitating configurations. When these systems collapse in a bosenova, part of their mass is turned into relativistic radiation that escapes the galaxy. We bound the mass that a single object can eject and sum it over realistic models of the Milky Way and of other types of galaxies. For gravitational atoms, with realistic stellar populations and stellar evolution, the stars of the Milky Way deplete at most $\epsGA_\star\simeq3\times10^{-9}$ of its halo, and the pump is equally negligible in other galaxies. The central black hole, Sgr~A$^*$, could reach $\epsGA\sim10^{-3}$, but only for a boson mass disfavoured by ultrafaint dwarf galaxies and beyond the validity of the perturbative treatment, where only a geometric bound on the dark matter flux is available. For boson stars feeding on the rest of the halo, these would radiate at most $\sim3\times10^{-5}$ of the Milky-Way halo and $0.1\%$ of a DF2-like one.
\end{abstract}

\maketitle

\section{Introduction}
\label{sec:intro}

Ultralight bosons (ULBs) with masses $10^{-22}\eV\lesssim m\lesssim 1\eV$ are among the most studied alternatives to weakly interacting massive particles as the constituents of dark matter (DM) \cite{Hui:2016ltb}. The prototypical example is the QCD axion \cite{Peccei:1977hh,Weinberg:1977ma,Wilczek:1977pj}, which lies in the higher-mass range of the spectrum,  together with its axion-like generalisations, which arise abundantly in string compactifications \cite{Arvanitaki:2009fg} and produce a plethora of very low mass axion-like particles. In galactic halos their occupation numbers are enormous, so the field behaves as a classical wave on the de Broglie scale $\lambda_{\rm dB}=2\pi/(m\vdm)$, where $\vdm$ is the velocity of the DM. Moreover, any axion-like potential $V(\phi)=-m^2f_a^2\cos(\phi/f_a)$ also implies an attractive quartic self-interaction, $\lambda=-m^2/f_a^2$. The coupling is tiny, but the large occupation numbers compensate for it, and it can drive the formation of bound configurations formed solely by accretion \cite{Levkov:2018kau}.

The capture of ULB DM by a star or a compact object of mass $M$ creates bound states described by a Schr\"odinger--Newton system of equations, which are analogous to  those of a hydrogen atom, with gravitational coupling $\alpha=Mm$ and ``Bohr radius'' $\RGA=1/(Mm^2)$, hence the name \emph{gravitational atom} (GA) \cite{Banerjee:2019epw,Baumann:2018vus}. In \cite{Budker:2023sex} it has been shown that self-interactions populate these states from the unbound DM waves. When the atom is smaller than the de Broglie wavelength of the DM (a \emph{dense} atom), Bose enhancement makes its mass grow exponentially, on the self-interaction relaxation time, until the attractive self-interaction overcomes the gradient pressure. The atom then collapses  \cite{Levkov:2016rkk,Eby:2016cnq}, and radiates an $\mathcal O(1)$ fraction of its mass in relativistic bosons: a \emph{bosenova}. The remnant can grow again, and the cycle should in principle repeat itself through cosmic evolution. This could in principle constitute a kind of ``DM pump'' which could potentially deplete part of the DM galactic halo. 

Gravity and self-interactions also act on the galactic DM halo itself. Kinetic relaxation populates the low-momentum modes of a virialised halo until a coherent, self-gravitating condensate nucleates \cite{Levkov:2018kau,Jain:2023tsr}: a \emph{boson star} (BS) \cite{Liebling:2012fv}. BSs can also be inherited from the early Universe, for instance from axion miniclusters \cite{Fox:2023xgx,Maseizik:2024qya}. Embedded in the halo, a star keeps growing by Bose-enhanced scattering of the gas, as a dense atom does. Gravity makes it grow if its virial temperature is not much below that of the gas, and evaporate otherwise \cite{Chan:2022bkz,Wang:2025vhy}, while the self-interaction adds a growth rate that dominates for the decay constants of interest here \cite{Wang:2025vhy}, just as the atomic case. With an attractive coupling, however, a star is stable only up to a maximum mass, $\McBS\simeq10f_a/m$ \cite{Chavanis:2011zi,Eby:2016cnq,Schiappacasse:2017ham}; we take the mass and radius of a star from the variational ansatz of Ref.~\cite{Schiappacasse:2017ham}. A heavier star collapses and radiates a fraction of its mass in relativistic bosons \cite{Levkov:2016rkk}, after which the remnant grows again. BSs can therefore constitute a second kind of DM pump, fed directly by the halo rather than by the stars of the galaxy.

This picture suggests a simple question. A galaxy contains $\sim10^{11}$ stars and compact remnants embedded in a DM halo. If each of them hosts an atom that repeatedly turns cold DM into radiation that escapes the galaxy, could these processes deplete a significant fraction of the halo over cosmic time? A positive answer would convert DM into radiation without any coupling to the Standard Model, and could produce DM-deficient galaxies such as NGC~1052-DF2 \cite{vanDokkum:2018vup} or NGC~1277 \cite{Comeron:2023}. Moreover, it could serve to impose constraints on axion-like self-interactions, since the depletion should not be in tension with the current DM distributions observed. On the other hand, a negative answer could bound the fraction of DM processed by stars.

Here we carry out this calculation. We derive an upper bound on the mass ejected by a single object. We arrive at the conclusion that bosenovae caused by the collapse of GAs formed around stars can deplete at most $\epsGA_\star\simeq3\times10^{-9}$ of the halo for a Milky Way-type galaxy. The central black hole (BH), Sgr~A$^*$, does better in a separate window of boson masses, but through a less controlled estimate and for masses disfavoured by other observations. For the case of BSs formed in the halo, these can remove at most $\sim3\times10^{-5}$ of the Milky-Way halo. No channel comes close to emptying a galaxy, but the bursts of the Sun's own atom could be detectable. We also estimate how these numbers could change in other galaxies. 

Section~\ref{sec:models} presents the theoretical framework and the constraints common to both channels. Section~\ref{sec:GAs} studies the bosenovae of GAs: the bound for a single object, the Galactic model, the results for the Milky Way, including Sgr~A$^*$, and for other galaxies and their central BHs. Section~\ref{sec:bosestars} studies BSs, and Sec.~\ref{sec:conclusions} compares all the channels and discusses how the bursts could be detected. We use natural units $\hbar=c=G=1$, so that masses are measured in units of the Planck mass $M_{\rm Pl}=G^{-1/2}=1.22\times10^{19}\GeV$, and quote numerical values in conventional units.

\section{Theoretical framework}
\label{sec:models}

All the processes studied in this paper follow from the same non-relativistic field theory. We first derive it and then summarise the two kinds of bound configurations that it predicts: GAs, held by the potential of a star or a compact object (Sec.~\ref{sec:GA}), and BSs, held by their own gravity (Sec.~\ref{sec:BSmodel}), together with the constraints common to both.

We consider a real scalar $\phi$ minimally coupled to gravity, with Lagrangian density
\begin{equation}
\mathcal L=\tfrac12\,g^{\mu\nu}\partial_\mu\phi\,\partial_\nu\phi-V(\phi)
\label{eq:L}
\end{equation}
and potential
\begin{equation}
V(\phi)=m^2f_a^2\big[1-\cos(\phi/f_a)\big]\simeq\tfrac12m^2\phi^2+\tfrac{\lambda}{4!}\phi^4,
\label{eq:V}
\end{equation}
with $\lambda=-m^2/f_a^2$. Two approximations reduce Eqs.~\eqref{eq:L} and~\eqref{eq:V} to a non-relativistic wave equation. First, the occupation numbers of the DM modes, are enormous, so the field can be treated as a classical wave \cite{Hui:2016ltb}. Second, the DM is non-relativistic, $v\ll1$, and the gravitational fields are weak, $|\Phi|\ll1$. In the Newtonian metric $ds^2=(1+2\Phi)\,dt^2-(1-2\Phi)\,d\mathbf x^2$ we write
\begin{equation}
\phi=\frac{1}{\sqrt{2m}}\big(\psi\,e^{-imt}+\psi^*e^{imt}\big),
\label{eq:NR}
\end{equation}
with $\psi$ slowly varying, $|\partial_t\psi|\ll m|\psi|$, and average over the fast oscillations. To leading order in $v^2$, $\Phi$ and $|\lambda|\phi^2/m^2$, the field obeys the Gross--Pitaevskii--Poisson system
\begin{align}
i\partial_t\psi&=-\frac{\nabla^2\psi}{2m}+m\Phi\,\psi+g|\psi|^2\psi,\label{eq:GP}\\
\nabla^2\Phi&=4\pi\big(\rho_{\rm b}+m|\psi|^2\big),\label{eq:Poisson}
\end{align}
with $g=\lambda/(8m^2)=-1/(8f_a^2)$, DM density $\rhodm=m|\psi|^2$ and $\rho_{\rm b}$ the density of baryons and compact objects. It breaks down only inside a collapsing configuration, where the density approaches $m^2f_a^2$; the relativistic emission of a bosenova lies beyond it.

\paragraph{Kinetic relaxation.} In a halo the field is a superposition of waves with random phases and a nearly Maxwellian distribution of momenta. The interactions in Eqs.~\eqref{eq:GP}--\eqref{eq:Poisson} redistribute the occupation numbers on two time scales: the self-interaction relaxation time \cite{Budker:2023sex}
\begin{equation}
\taurel=\frac{64\,m^{7}\vdm^{2}}{\lambda^{2}\rhodm^{2}}=\frac{64\,m^{3}f_a^{4}\vdm^{2}}{\rhodm^{2}},
\label{eq:taurel}
\end{equation}
where $\vdm$ is the typical velocity of the DM, and the gravitational one \cite{Levkov:2018kau}
\begin{equation}
\tau_{\rm gr}=\frac{b\sqrt2\,m^3\vdm^6}{12\pi^3\rhodm^2\Lambda},
\label{eq:taugr}
\end{equation}
with $b\simeq0.7$ and $\Lambda=\ln(m\vdm L)$ for a system of size $L$. Both scale as $\rhodm^{-2}$. Relaxation feeds every bound configuration in this paper: it populates the bound states of the potential of a star (Sec.~\ref{sec:GA}) and the low-momentum modes of the gas itself (Sec.~\ref{sec:BSmodel}). In both cases the mechanism is the same. Two-to-two scatterings $k_1+k_2\to k_3+b$ transfer particles from the gas into a bound state $b$, and the inverse process removes them. The first gives a capture rate that does not depend on the occupation of $b$; once $b$ is populated, Bose enhancement makes both processes proportional to its occupation, so that the configuration grows exponentially if stimulated capture beats the inverse process and is depleted otherwise. For a Maxwellian gas of temperature $T_g$ the gas feeds a bound configuration efficiently only if its binding energy is comparable to the temperature of the gas, and depletes it otherwise. This is what separates dense from dilute atoms (Sec.~\ref{sec:GA}) and growing from evaporating stars (Sec.~\ref{sec:BSmodel}).

\subsection{Gravitational atoms}
\label{sec:GA}

Near a star or a compact object of mass $M$ the potential $\Phi=-M/r$ dominates, and Eq.~\eqref{eq:GP} becomes
\begin{equation}
\Big(i\partial_t+\frac{\nabla^2}{2m}+\frac{\alpha}{r}\Big)\psi=g|\psi|^2\psi,
\label{eq:GPatom}
\end{equation}
with $\alpha=Mm$. In the absence of self-interactions the bound states are hydrogenic, $\psi_{nlm}$, with ground state $\psi_{100}=e^{-r/\RGA}/\sqrt{\pi(\RGA)^3}$ and
\begin{equation}
\RGA=\frac{1}{Mm^2}.
\label{eq:Rstar}
\end{equation}
The scattering states are Coulomb waves, and the parameter that controls how strongly the incoming DM waves of velocity $\vdm$ are focused by the object is
\begin{equation}
\xifoc\equiv\frac{2\pi\alpha}{\vdm}=\frac{2\pi Mm}{\vdm}.
\label{eq:xi}
\end{equation}

In \cite{Budker:2023sex} it was shown that the self-interaction populates these bound states through the processes of Sec.~\ref{sec:models}: two-to-two scatterings $k_1+k_2\to k_3+(nlm)$ capture particles at a rate $C$, and once the ground state is populated, Bose enhancement makes stimulated capture compete with stripping, so that the bound mass obeys
\begin{equation}
\dot M^{\rm GA}=C+\GGA\,\MGA.
\label{eq:growth}
\end{equation}
They derived the rates by treating $g$ perturbatively, for DM waves with a Maxwellian distribution centred at velocity $\vdm$ with dispersion $\sigma=\vdm/\sqrt2$. The rates are controlled by the relaxation time $\taurel$ of Eq.~\eqref{eq:taurel}, and the net rate between stimulated capture and stripping of bosons in the fundamental state has the limits \cite{Budker:2023sex}
\begin{equation}
\GGA\simeq\begin{cases}
+\dfrac{1}{0.3\,\taurel}, & \xifoc\gg1 \quad\text{(dense)},\\[2ex]
-\dfrac{\xifoc^{4}}{10\,\taurel}, & \xifoc\ll1 \quad\text{(dilute)},
\end{cases}
\label{eq:Gamma}
\end{equation}
with the sign change at $\xifoc\simeq1$. In the dilute regime the bound mass saturates at the small value $C/|\GGA|$, which fixes the capture coefficient there, $C_{\rm dil}=0.04\,\xifoc^{7}\rhodm(\RGA)^{3}/(\pi^{3/2}\taurel)$ \cite{Budker:2023sex}. In the dense regime the mass grows linearly as $Ct$ for $t\lesssim1/\GGA$ and exponentially afterwards. Since Ref.~\cite{Budker:2023sex} gives no closed form for its normalisation, we write
\begin{equation}
C=\frac{\rhodm(\RGA)^{3}}{\pi^{3/2}\taurel}\,\min\Big\{0.04\,\xifoc^{7},\ \,\xifoc^{3}\Big\},
\label{eq:C}
\end{equation}
i.e.\ we use the dilute result until it crosses the dense asymptote. This choice barely affects our results, because the stimulated term dominates where the ejection is largest (Sec.~\ref{sec:results}).

We assume that the growth stops when the central density reaches
\begin{equation}
\rhocGA\simeq\frac{16\alpha^{2}m^{4}}{|\lambda|}=16\,\alpha^{2}m^{2}f_a^{2},
\label{eq:rhocrit}
\end{equation}
at which the attractive self-interaction can no longer be balanced by the gradient pressure. For the ground state, so the critical mass of the atom is
\begin{equation}
\McGA=\pi(\RGA)^{3}\rhocGA=\frac{16\pi f_a^{2}}{Mm^{2}}.
\label{eq:Mc}
\end{equation}
Solving Eq.~\eqref{eq:growth} from an empty atom, $\MGA(t)=(C/\GGA)(e^{\GGA t}-1)$, the time of the first bosenova is
\begin{equation}
\tcGA=\frac{1}{\GGA}\ln\Big(1+\frac{\GGA\McGA}{C}\Big),
\label{eq:tc}
\end{equation}
which reduces to $\McGA/C$ when capture dominates ($\GGA\McGA\ll C$) and to $0.3\,\taurel\ln(\GGA\McGA/C)$ when the exponential phase does.

\begin{figure}[t]
\centering
\includegraphics[width=\columnwidth]{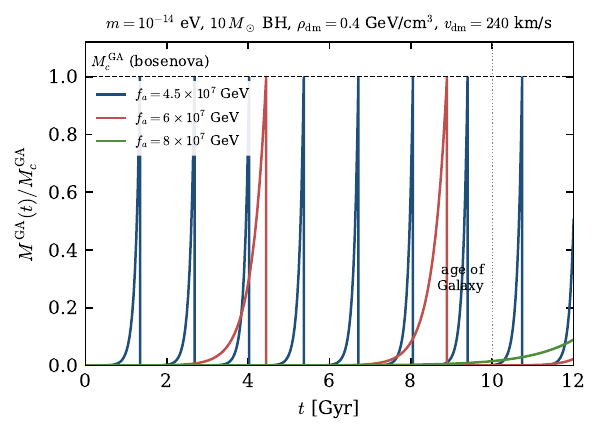}
\caption{Growth of a GA around a $10\Msun$ BH at the solar position, Eq.~\eqref{eq:growth}, for $m=10^{-14}\eV$ and three values of $f_a$. For these parameters the atom is dense ($\xifoc\simeq6$) and grows exponentially by stimulated capture until it reaches the critical mass $\McGA$ (dashed line) and explodes; with $\dGA=1$ it restarts from zero and the cycle repeats every $\tcGA$, Eq.~\eqref{eq:tc}, which scales as $f_a^{4}$ through $\taurel$. Within the age of the Galaxy (dotted line) the atom explodes seven times for $f_a=4.5\times10^{7}\GeV$, twice for $6\times10^{7}\GeV$ and never for $8\times10^{7}\GeV$.}
\label{fig:growth}
\end{figure}

In Fig.~\ref{fig:growth} we show the mass of an atom around a $10\Msun$ BH at the solar position for $m=10^{-14}\eV$ and three decay constants, assuming that each bosenova resets the atom to zero mass. For these parameters the atom is dense and the stimulated term dominates, so it grows exponentially from the first capture and goes through a bosenova every $1.3\Gyr$ for $f_a=4.5\times10^{7}\GeV$. Since $\tcGA\propto\taurel\propto f_a^{4}$, a slightly larger $f_a$ already stretches the cycle to $4.4\Gyr$, and at $f_a=8\times10^{7}\GeV$ the atom does not reach $\McGA$ within the age of the Galaxy.

What happens after the first bosenova depends on details of the collapse that are not fully under control. Simulations of collapsing BSs find that a fraction $\dGA\sim0.1$--$1$ of the mass is radiated in relativistic particles, while the rest remains as a subcritical remnant \cite{Levkov:2016rkk,Eby:2016cnq}. For $\dGA=1$ the atom restarts from zero after each explosion; for $\dGA<1$ the remnant regrows in the shorter time
\begin{equation}
\trepGA=\tcGA-\frac{1}{\GGA}\ln\Big[1+(1-\dGA)\frac{\GGA\McGA}{C}\Big],
\label{eq:trep}
\end{equation}
and each explosion ejects only $\dGA\McGA$. 

\subsection{Boson stars}
\label{sec:BSmodel}

Kinetic relaxation also acts on the gas itself, and in principle it should populate the low-momentum modes of a virialized gas until a coherent condensate nucleates. If its own gravity can hold it, the condensate settles into a BS (an axion star, for an axion-like field), a self-gravitating ground state of Eqs.~\eqref{eq:GP}--\eqref{eq:Poisson}. BSs can also be inherited from earlier epochs, for instance from axion miniclusters \cite{Fox:2023xgx,Maseizik:2024qya,Wang:2025vhy}.

Far from any star, $\rho_{\rm b}=0$ and the potential is sourced by the field itself, $\Phi(\bm x)=-m\int d^3x'\,|\psi(\bm x')|^2/|\bm x-\bm x'|$, so that Eq.~\eqref{eq:GP} becomes
\begin{equation}
\Big(i\partial_t+\frac{\nabla^2}{2m}+m^2\!\int d^3x'\,\frac{|\psi(\bm x')|^2}{|\bm x-\bm x'|}\Big)\psi=g|\psi|^2\psi,
\label{eq:GPstar}
\end{equation}
the analogue of Eq.~\eqref{eq:GPatom} with the Coulomb potential of the object replaced by the self-gravity of the condensate.

We describe a star of mass $\MBS$ by the ansatz $\psi\propto{\rm sech}(r/\RBS)$ of Ref.~\cite{Schiappacasse:2017ham}. For $\lambda=-m^2/f_a^2$ the only mass and length scales of Eq.~\eqref{eq:GPstar} are $f_a/m$ and $1/(mf_a)$, and in the corresponding units,
\begin{equation}
\tilde N=\frac{\MBS m}{f_a},\qquad \tilde R=\RBS mf_a,
\label{eq:units}
\end{equation}
the energy of the star is the sum of its gradient, gravitational and self-interaction terms,
\begin{equation}
E=\frac{f_a^3}{m}\Big(\frac{c_1\tilde N}{\tilde R^2}-\frac{c_2\tilde N^2}{\tilde R}-\frac{c_3\tilde N^2}{\tilde R^3}\Big),
\label{eq:Evar}
\end{equation}
with $c_1=(12+\pi^2)/(6\pi^2)\simeq0.369$, $c_2=6[12\zeta(3)-\pi^2]/\pi^4\simeq0.281$ and $c_3=(\pi^2-6)/(8\pi^5)\simeq1.58\times10^{-3}$. Equilibria satisfy $\partial E/\partial\RBS=0$, whose stable solution is
\begin{equation}
\tilde R=\frac{c_1+\sqrt{c_1^2-3c_2c_3\tilde N^2}}{c_2\tilde N}.
\label{eq:Rstable}
\end{equation}
Lighter stars are larger and more dilute. We measure the size of a star by its half-mass radius, $r_h\simeq1.49\,\RBS$.

A star embedded in a gas grows by Bose-enhanced scattering of the gas, mediated by gravity and by the self-interaction. Since the star itself seeds the stimulated scattering, its mass obeys Eq.~\eqref{eq:growth} without the capture term,
\begin{equation}
\dot M^{\rm BS}=\GBS\MBS,
\label{eq:Mdot}
\end{equation}
which we cap at the geometric flux onto the star. The gravitational rate is \cite{Chan:2022bkz,Wang:2025vhy}
\begin{equation}
\Gamma_{\rm gr}=\frac{(4\pi)^2m^3\rhodm^2}{k_g^6}\,\gamma(k_gr_s),
\label{eq:Ggr}
\end{equation}
where $k_g=m\vdm/\sqrt{3/2}$ is the typical momentum of the gas, whose rms velocity we identify with $\vdm$, and $r_s\simeq0.88\,\RBS$ the half-density radius of the star. In the simulations of Ref.~\cite{Chan:2022bkz}, $\gamma\simeq5.8$ for $0.7\lesssim k_gr_s\lesssim2.5$, falls as $(k_gr_s)^4$ for heavier stars, and becomes negative, $\gamma\simeq-2.6\,(k_gr_s)^2$, for lighter ones, which evaporate. Self-interactions add \cite{Wang:2025vhy}
\begin{equation}
\Gamma_\lambda\simeq\frac{60\,\lambda^2\rhodm^2}{16\pi^2m^5k_g^2}\simeq\frac{36}{\taurel},
\label{eq:Glam}
\end{equation}
so that a condensate grows exponentially, as a dense atom does. Equation~\eqref{eq:Glam} was derived for stars no larger than the de Broglie wavelength of the gas \cite{Wang:2025vhy}. For much larger stars it overestimates the rate: such a star is locally a homogeneous condensate of chemical potential $\mu<0$, and by the balance argument of Sec.~\ref{sec:models} its net rate per particle is suppressed by $|\mu|/T_g\sim(k_gr_s)^{-2}$, as the gravitational one is. For stars much larger than $\lambda_{\rm dB}$, Eq.~\eqref{eq:Glam} is therefore an upper estimate. We take the total rate of the star to be the sum of the two,
\begin{equation}
\GBS=\Gamma_{\rm gr}+\Gamma_\lambda.
\label{eq:rate}
\end{equation}
Both come from Bose-enhanced two-to-two scatterings, mediated by gravity and by the contact interaction, so the sum neglects their interference, which is known to speed up the relaxation of an attractive gas \cite{Jain:2023tsr}. The interference is at most of order $2\sqrt{|\Gamma_{\rm gr}|\Gamma_\lambda}$. At the maxima of Sec.~\ref{sec:bosestars} $\Gamma_\lambda$ exceeds $|\Gamma_{\rm gr}|$ by more than a factor of three along the whole cycle, so the interference changes the results by less than $10\%$ in the Milky Way, and by up to a factor of order unity in dwarf galaxies.

The growth stops at a maximum mass, the analogue of the critical mass of an atom. Equation~\eqref{eq:Rstable} has a solution only for $\tilde N\le\tilde N_c=c_1/\sqrt{3c_2c_3}\simeq10.1$: with an attractive coupling, BSs are stable only up to \cite{Chavanis:2011zi,Eby:2016cnq,Schiappacasse:2017ham}
\begin{equation}
\McBS\simeq10\,\frac{f_a}{m}.
\label{eq:Mmax}
\end{equation}
A star is a distinct object only if it is denser than the gas, which along the whole cycle requires $f_a^4\gtrsim\rhodm/(8m^2)$.

Only $\Gamma_{\rm gr}$ and the cap depend on the mass of the star, and, as for an atom, we evaluate both at the critical mass $\McBS$. At the maxima of Sec.~\ref{sec:bosestars} this overestimates the rate averaged over a cycle by less than $25\%$. With $\GBS$ constant, Eq.~\eqref{eq:Mdot} gives, from an initial mass $M_0^{\rm BS}$, $\MBS(t)=M_0^{\rm BS}e^{\GBS t}$, and the star reaches $\McBS$ at
\begin{equation}
\tcBS=\frac{1}{\GBS}\ln\frac{\McBS}{M_0^{\rm BS}}.
\label{eq:tcBS}
\end{equation}
Unlike an atom, a star does not start empty, so $\tcBS$ depends on the initial mass, which is set by the formation of the star; and the star never reaches $\McBS$ if it evaporates, $\GBS<0$.

A star that grows beyond $\McBS$ collapses. The collapse proceeds through a series of bursts that radiate a fraction $\dBS\simeq0.3$ of its mass in mildly relativistic bosons \cite{Levkov:2016rkk}. The remnant, $(1-\dBS)\McBS$, grows again. In an unlimited supply of gas a star therefore cycles between $(1-\dBS)\McBS$ and $\McBS$. Each cycle lasts the time the remnant takes to regrow,
\begin{equation}
\trepBS=\frac{1}{\GBS}\ln\frac{1}{1-\dBS},
\label{eq:tcyc}
\end{equation}
and ends with the emission of $\dBS \McBS$.

\begin{figure}[t]
\centering
\includegraphics[width=\columnwidth]{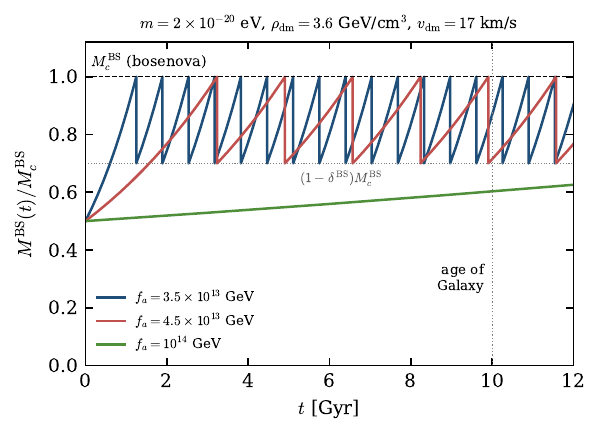}
\caption{Mass of a BS in the core of a Fornax-like dwarf spheroidal ($r=0.2\kpc$), starting from $\McBS/2$, for three decay constants. Each time the star reaches $\McBS$ (dashed line) it collapses and radiates a fraction $\dBS=0.3$ of its mass.}
\label{fig:bsgrowth}
\end{figure}

In Fig.~\ref{fig:bsgrowth} we show the mass of a BS in the core of a dwarf spheroidal for three decay constants. Just above the smallest $f_a$ at which the remnant is denser than the gas, the self-interaction makes the star grow quickly, and it goes through a bosenova every $0.6\Gyr$. Since $\Gamma_\lambda\propto f_a^{-4}$, a slightly larger $f_a$ already stretches the cycle to $1.7\Gyr$, and at still larger $f_a$, where gravity takes over, the star grows too slowly to reach $\McBS$ within the age of the Galaxy.

GAs and BSs feed on the same ambient DM, and two constraints apply to both. First, the ambient DM must be a smooth medium. The attractive self-interaction gives it a negative contribution to the squared sound speed, $g\rhodm/m^2=-\rhodm/(8m^2f_a^2)$, which must be smaller in magnitude than the squared velocity dispersion, $\sigma^2=\vdm^2/2$:
\begin{equation}
\rhodm<4m^2\vdm^2f_a^2.
\label{eq:stab}
\end{equation}
Where it fails, the DM is strongly self-interacting and requires a different treatment.

Second, attractive self-interactions modify the growth of linear perturbations. In the equation for the DM overdensity the term $(k/k_\lambda)^2$ enhances the growth of modes with comoving momentum above $k_\lambda(a)=a\,(16\pi m^4/|\lambda|)^{1/2}$ \cite{Budker:2023sex}. Requiring that this effect is negligible at the largest momentum where the linear power spectrum is measured, $k\simeq1\,{\rm Mpc}^{-1}$, at matter-radiation equality gives the conservative bound
\begin{equation}
f_a\gtrsim3.7\times10^{6}\GeV\,\Big[\frac{10^{-14}\eV}{m}\Big],
\label{eq:cosmo}
\end{equation}
which we display as a shaded region in all figures. We also impose the Lyman-$\alpha$ bound on the boson mass, $m\gtrsim2\times10^{-20}\eV$ \cite{Rogers:2020ltq}, and do not consider the regions excluded by either bound further.

The decay constants that matter below are also far smaller than those for which vacuum misalignment produces all the DM, $f_a\gtrsim6\times10^{14}\GeV$ for $m=10^{-15}\eV$ \cite{Hui:2016ltb}. The DM must then be produced by another mechanism, such as kinetic misalignment \cite{Co:2019jts}, and we take its abundance as given. Restricting the analysis to the misalignment line would only strengthen our conclusions, since all the rates fall with $f_a$.

\section{Gravitational atom pumps}
\label{sec:GAs}

A simple bound follows from mass conservation. Let $\MGA(t)$ be the mass of the atom and
\begin{equation}
\dot M_{\rm in}(\MGA)=\min\{C+\GGA\MGA,\ \Fgeo\}
\label{eq:Min}
\end{equation}
the rate at which it takes DM from the halo: the net perturbative rate of Eq.~\eqref{eq:growth}, capped by the geometric flux $\Fgeo$ through the atom, defined below. Between bosenovae the mass of the atom changes only through $\dot M_{\rm in}$, and in a bosenova it only decreases, so the mass radiated up to a time $T$ is the mass captured minus the mass still bound at $T$ (the atom starts empty),
\begin{equation}
M_{\rm ej}(T)=\int_0^T\dot M_{\rm in}\,dt-\MGA(T).
\label{eq:conservation}
\end{equation}
We bound the two terms separately, splitting the integral at the first bosenova, $\tcGA$. Before it nothing has been ejected, so the captured mass is the mass of the atom, which by definition of $\tcGA$ is the critical mass:
\begin{equation}
\int_0^{\tcGA}\dot M_{\rm in}\,dt=\MGA(\tcGA)=\McGA.
\label{eq:step1}
\end{equation}
After it the atom explodes whenever it reaches $\McGA$, so $\MGA(t)\le\McGA$ at all times, and since $\dot M_{\rm in}$ grows with $\MGA$ (the atom is dense, $\GGA>0$), the capture rate never exceeds that of an atom at the critical mass:
\begin{equation}
\int_{\tcGA}^T\dot M_{\rm in}\,dt\le(T-\tcGA)\min\{C+\GGA\McGA,\Fgeo\}.
\label{eq:step2}
\end{equation}
Finally, each bosenova leaves a remnant of at least $(1-\dGA)\McGA$ (Sec.~\ref{sec:GA}), and the atom only grows between bosenovae, so after $\tcGA$ it never falls below that remnant:
\begin{equation}
\MGA(T)\ge(1-\dGA)\McGA.
\label{eq:step3}
\end{equation}
Inserting Eqs.~\eqref{eq:step1}--\eqref{eq:step3} into Eq.~\eqref{eq:conservation} gives, for $T>\tcGA$, $M_{\rm ej}(T)\le\MejGA$ with
\begin{equation}
\begin{split}
\MejGA={}&\underbrace{\dGA\McGA}_{\rm store}\\
&+\underbrace{(T-\tcGA)\min\{C+\GGA\McGA,\Fgeo\}}_{\rm pump}
\end{split}
\label{eq:bound}
\end{equation}
and $M_{\rm ej}=0$ if $\tcGA>T$. The two terms are:
\begin{itemize}
\item \emph{Store}: the part of the mass of the atom at its first bosenova, captured before $\tcGA$, that is radiated and not replaced; it is counted at most once. We take its largest value, $\dGA=1$: at the maxima below it is less than $10^{-6}$ of the ejected mass.
\item \emph{Pump}: the DM captured after $\tcGA$, at the throughput $C+\GGA\McGA$ of an atom at $\McGA$, capped by $\Fgeo=\rhodm\vdm\,\pi(\RGA)^2\,(1+\xifoc^2/2\pi^2)$ . This is the flux of DM through the gravitationally focused cross section of the atom: no scheme can capture more DM than flows through it, and where the perturbative rate exceeds it the perturbative treatment has broken down anyway. 
\end{itemize}
The pump term is saturated for small $\dGA$, where the atom sits at $\MGA\simeq\McGA$ and steadily converts the captured DM into radiation with throughput $C+\GGA\McGA$; for $\dGA=1$ the ejection is smaller by a logarithm. Using Eqs.~\eqref{eq:taurel}--\eqref{eq:Mc}, the two terms of the throughput are
\begin{align}
\GGA\McGA&=\frac{5\pi}{6}\frac{\rhodm^{2}}{M\,m^{5}f_a^{2}\vdm^{2}},\label{eq:GMc}\\
C_{\rm dense}&=\frac{\pi^{3/2}}{8}\frac{\rhodm^{3}}{m^{6}f_a^{4}\vdm^{5}},\label{eq:Cnum}
\end{align}
where $C_{\rm dense}$ is the dense asymptote of Eq.~\eqref{eq:C}.

Besides Eqs.~\eqref{eq:stab} and~\eqref{eq:cosmo} and the Lyman-$\alpha$ bound, which apply to both channels, the formation mechanism of Ref.~\cite{Budker:2023sex} rests on assumptions that translate into conditions on the object and its environment. We impose all of the following:
\begin{enumerate}
\item[(i)] \emph{Point-like source}: $\RGA>R_{\rm obj}$, the radius of the object. This matters only for stars, not for compact remnants.
\item[(ii)] \emph{Non-relativistic bound state}: $\alpha<1/2$, and $\McGA<M$ so that the self-gravity of the atom is negligible.
\item[(iii)] \emph{Perturbativity of the background}: the self-interaction energy of the ambient waves must be smaller than the gravitational energy at $\RGA$, $|g||\psi_w|^2<mM/\RGA$, i.e.\ $\rhodm<\rhocGA/2$. Otherwise the ambient DM is not governed by the object at all.
\item[(iv)] \emph{Dense regime}: $\xifoc>1$, so that $\GGA>0$; dilute atoms never reach $\McGA$ once (iii) is imposed.
\item[(v)] \emph{Time}: $\tcGA$ shorter than the time the object has been available, which for a main-sequence star is its lifetime.
\end{enumerate}
Condition (iii) has the largest impact. Since $\rhocGA\propto M^2m^2f_a^2$, at small $f_a$ or small $m$ only the objects in the low-density outskirts of the halo can host an atom, and the steep growth of the throughput at small $f_a$ is cut off. Like Eq.~\eqref{eq:stab}, it limits the applicability of the calculation, not the parameters: where it fails, the DM is strongly self-interacting and requires a different treatment.

\begin{figure}[t]
\centering
\includegraphics[width=\columnwidth]{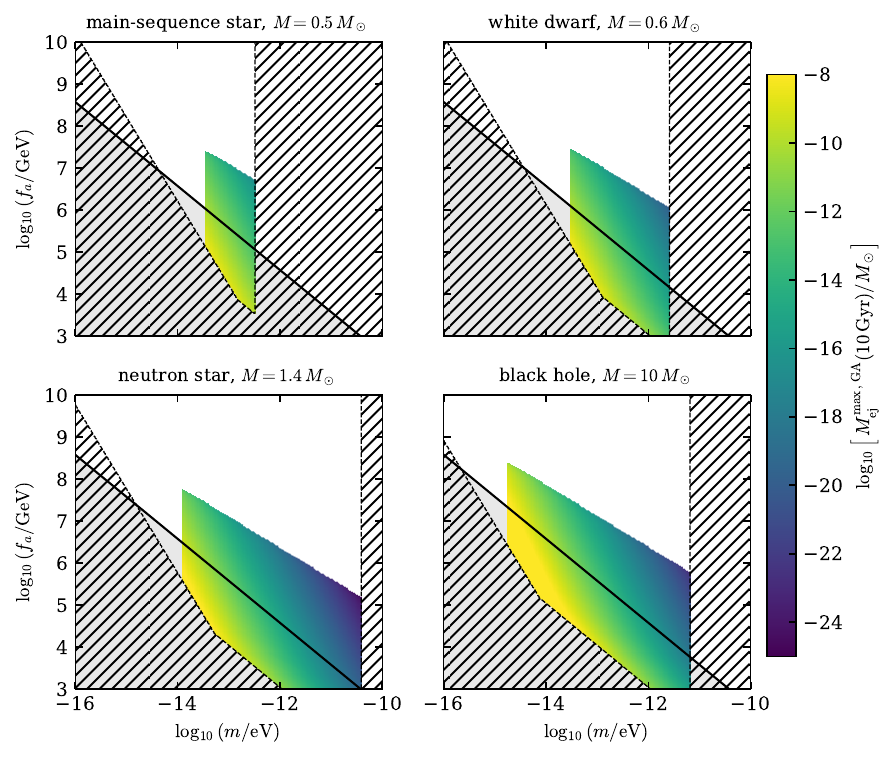}
\caption{Maximal mass ejected in $10\Gyr$, Eq.~\eqref{eq:bound}, by a single object located at the solar position ($\rhodm=0.4\,{\rm GeV/cm^3}$, $\vdm=240\kms$), for four representative objects. Hatched: regions where Eq.~\eqref{eq:stab} or one of the conditions (i)--(iv) of Sec.~\ref{sec:GAs} fails (the dashed line delimits them); white: valid but $\tcGA>10\Gyr$. The grey region is excluded by linear structure formation, Eq.~\eqref{eq:cosmo}.}
\label{fig:objects}
\end{figure}

In Fig.~\ref{fig:objects} we show the bound of Eq.~\eqref{eq:bound} for four objects at the solar position. The allowed region is limited from below by condition (iii) and Eq.~\eqref{eq:stab}, from the left by the dense condition, $m>\vdm/(2\pi M)$, and from above by the requirement $\tcGA<10\Gyr$. Even in its most favourable corner, at the edge of the dense regime, a single object ejects less than $3\times10^{-8}\Msun$ over the age of the Galaxy. BHs fare best, because they are dense down to smaller $m$, where the throughput, $\propto m^{-5}$, is largest.

\subsection{Galactic model}
\label{sec:galaxy}

The total ejected mass is obtained by summing Eq.~\eqref{eq:bound} over the objects in the Galaxy. Besides the smooth stellar population, the sum includes a single point object representing the central supermassive BH, Sgr~A$^*$ (Sec.~\ref{sec:sgra}), so that
\begin{align}
\epsGA\equiv\frac{M_{\rm ej}^{\rm tot}}{M_{\rm dm}}=\frac{1}{M_{\rm dm}}\int dM_i\,\frac{dN}{dM_i}\int d^{3}x\,n_\star(\bm x)\nonumber\\
\times\sum_{\rm phases}\MejGA\big(M,R_{\rm obj},\tau;\rhodm(\bm x),\vdm(\bm x)\big)\nonumber\\{}+\MejGA\big(M_{\rm BH}^{\rm Sgr\,A^*},R_{\rm BH}^{\rm Sgr\,A^*},T;\nonumber\\
\rhodm(\bm x^{\rm Sgr\,A^*}),\vdm(\bm x^{\rm Sgr\,A^*})\big),
\label{eq:eps}
\end{align}
where $M_{\rm ej}^{\rm tot}$ is the mass ejected by all the objects of the Galaxy over its age and $M_{\rm dm}=1.2\times10^{12}\Msun$ the total mass of its halo; $M_i$ is the initial mass of a star, $dN/dM_i$ the number of stars formed per unit initial mass and $n_\star(\bm x)$ their spatial distribution, normalised to unity, summed over the components of the stellar population. $\MejGA$ is the bound of Eq.~\eqref{eq:bound}, evaluated for each object with its mass $M$, its radius $R_{\rm obj}$ and its duration $\tau$ in place of $T$, and with the DM density $\rhodm(\bm x)$ and velocity $\vdm(\bm x)$ at its position; the sum runs over the two phases of a star, main sequence and remnant, each with its own $M$, $R_{\rm obj}$ and $\tau$. The last term is Sgr~A$^*$, a single object of mass $M_{\rm BH}^{\rm Sgr\,A^*}=4\times10^{6}\Msun$ and Schwarzschild radius $R_{\rm BH}^{\rm Sgr\,A^*}=2M_{\rm BH}^{\rm Sgr\,A^*}$ that lasts $T=10\Gyr$ at the Galactic centre $\bm x^{\rm Sgr\,A^*}$, where the DM is evaluated at the radius of its atom, $\sim1\pc$ (Sec.~\ref{sec:sgra}).

\paragraph{Dark matter.} The halo is an NFW profile \cite{Navarro:1996gj} with the parameters of the McMillan fit \cite{McMillan:2016jtx}: scale radius $19.6\kpc$ and $\rhodm=0.39\,{\rm GeV/cm^3}$ at the solar radius, $R_0=8.2\kpc$. The rates depend on the velocity of the DM relative to the object, which for a star on a nearly circular orbit is dominated by the circular speed $v_c$. We take $\vdm(r)=[v_c^2(r)+\sigma_0^2]^{1/2}$ with $\sigma_0=100\kms$, which gives $\vdm=240\kms$ at the Sun and stays above the stellar velocity dispersion in the bulge.

\begin{table}[t]
\caption{Components of the stellar population of the Milky Way: exponential thin and thick discs of scale length $R_d$ and scale height $z_d$, and a Hernquist bulge of scale radius $r_b$ (listed under $R_d$); formed stellar mass $M_\star$, metallicity $Z$ and formation history, Eq.~\eqref{eq:popdens}.}
\label{tab:components}
\footnotesize
\begin{ruledtabular}
\begin{tabular}{lccccl}
Component & $R_d$ & $z_d$ & $M_\star$ & $Z$ & Formation \\
 & [kpc] & [kpc] & [$\Msun$] & [$Z_\odot$] & \\
\hline
Thin disc & 2.6 & 0.3 & $6\times10^{10}$ & 1 & continuous, last $10\Gyr$ \\
Thick disc & 3.0 & 0.9 & $10^{10}$ & 0.25 & burst, $11\Gyr$ ago \\
Bulge & 0.5 & -- & $10^{10}$ & 1 & burst, $11\Gyr$ ago \\
\end{tabular}
\end{ruledtabular}
\end{table}

\paragraph{Stars.} The stellar population has three components: a thin disc that forms stars continuously over the last $10\Gyr$ (a simplification of the star-formation history of Ref.~\cite{Snaith:2014qxa}, from whose Fig.~7 we take the formation times and the division into a thick- and a thin-disc phase), and a thick disc and a Hernquist bulge \cite{Hernquist:1990be} that each form in a single burst $11\Gyr$ ago \cite{McMillan:2016jtx}. Their scale lengths, masses and metallicities are listed in Table~\ref{tab:components}. In each component the initial masses follow a Kroupa IMF \cite{Kroupa:2000iv}, so that the stars of component $j$ (formed mass $M_{\star,j}$) are distributed as
\begin{align}
\xi(M_i)&\propto\begin{cases}(M_i/0.5\Msun)^{-1.3}&(M_i<0.5\Msun)\\(M_i/0.5\Msun)^{-2.3}&(M_i\ge0.5\Msun)\end{cases}\nonumber\\
\frac{dN_j}{dM_i\,d^3x}&=\frac{M_{\star,j}\,\xi(M_i)}{\int_{0.08\Msun}^{100\Msun}\xi\,M\,dM}\,n_j(\bm x),\label{eq:popdens}\\
n_{\rm disc}&=\frac{e^{-R/R_d-|z|/z_d}}{4\pi R_d^2z_d},\nonumber\\
n_{\rm bulge}&=\frac{r_b}{2\pi r\,(r+r_b)^3},\nonumber
\end{align}
where $n_j$ is normalised to unity. The thin disc is sampled at six equally spaced formation times, and the thick disc and the bulge at $t_f=11\Gyr$. The present-day stellar mass, which enters $v_c$, is $70\%$ of the formed mass, the rest being returned to the interstellar medium \cite{Leitner:2010nh}.

\paragraph{Stellar evolution.} A star of initial mass $M_i$ and metallicity $Z$, formed a time $t_f$ ago, is treated as a main-sequence star for its lifetime $\tau_{\rm MS}(M_i,Z)$ and as a compact remnant afterwards. Eq.~\eqref{eq:bound} is evaluated separately for the two phases, over the time actually available to each. The main-sequence lifetime is the fit of Ref.~\cite{Hurley:2000pk} to the time to the base of the giant branch,
\begin{equation}
\tau_{\rm MS}(M_i,Z)=\frac{a_1+a_2x^{4}+a_3x^{5.5}+x^{7}}{a_4x^{2}+a_5x^{7}}\,\mathrm{Myr},
\label{eq:tms}
\end{equation}
with $x\equiv M_i/\Msun$ and coefficients $a_n$ that are cubic polynomials in $\log_{10}(Z/0.02)$, given in Ref.~\cite{Hurley:2000pk}. 

The remnant is a white dwarf (WD) for $M_i<8\Msun$, a neutron star (NS) for $8\le M_i<20\Msun$ and a BH above. WDs have mass $0.109M_i+0.394\Msun$ \cite{Kalirai:2008} and radius $0.01R_\odot(M/0.6\Msun)^{-1/3}$, and NSs have $1.4\Msun$ and $12$~km. BHs have $M_{\rm BH}=\max(0.5M_i,5\Msun)$, up to $50\Msun$. The wind mass loss of the progenitors, which lowers the masses of the BHs formed at solar metallicity \cite{Fryer:2011cx}, is not considered: capping $M_{\rm BH}$ at $20\Msun$ in the solar-metallicity populations changes the results of Secs.~\ref{sec:results} and~\ref{sec:galaxies} by less than $20\%$. NSs and BHs receive a natal kick, $265\kms$ for NSs, the dispersion of the Maxwellian adopted in population syntheses of Galactic NSs \cite{Sartore:2009wn}, and $265\kms\times1.4\Msun/M_{\rm BH}$ for BHs, which we add in quadrature to $\vdm$; the remnant is otherwise kept at the position of its progenitor. The fiducial model forms $9\times10^{9}$ WDs, $6\times10^{8}$ NSs and $2.4\times10^{8}$ BHs, consistent with population-synthesis estimates \cite{Sartore:2009wn}.

These analytic fits are crude, but they matter little here: the result is dominated by BH remnants, which last most of the age of the Galaxy, and it is only moderately sensitive to their masses.

\begin{figure}[tb]
\centering
\begin{tikzpicture}
\begin{axis}[
  width=\columnwidth, height=0.82\columnwidth,
  xmode=log, ymode=log, xmin=1e-3, xmax=15, ymin=0.08, ymax=100,
  xlabel={time since formation, $t$ [Gyr]},
  ylabel={initial stellar mass, $M_i$ [$M_\odot$]},
  tick align=inside, xtick pos=both, ytick pos=both,
  every axis label/.append style={font=\small},
  tick label style={font=\footnotesize},
  clip=true, axis on top, line width=0.4pt]
\definecolor{cMS}{HTML}{E8C547}
\definecolor{cGI}{HTML}{E08A1E}
\definecolor{cWD}{HTML}{1F4E79}
\definecolor{cNS}{HTML}{C0504D}
\definecolor{cBH}{HTML}{4F8F3A}
\fill[cMS!55] (0.001,0.08) -- (15,0.08) -- (15,0.9357) -- (7.762,1.112) -- (4.23,1.323) -- (2.417,1.572) -- (1.43,1.869) -- (0.8671,2.222) -- (0.5354,2.642) -- (0.336,3.141) -- (0.2143,3.735) -- (0.139,4.44) -- (0.09193,5.279) -- (0.0621,6.276) -- (0.04293,7.462) -- (0.03043,8.871) -- (0.02214,10.55) -- (0.01656,12.54) -- (0.01274,14.91) -- (0.01007,17.72) -- (0.008187,21.07) -- (0.006833,25.05) -- (0.00585,29.78) -- (0.005127,35.41) -- (0.004591,42.1) -- (0.00419,50.05) -- (0.003888,59.51) -- (0.00366,70.75) -- (0.003487,84.11) -- (0.003354,100) -- (0.001,100) -- cycle;
\fill[cWD!45] (15,0.9632) -- (7.297,1.168) -- (3.786,1.415) -- (2.071,1.716) -- (1.174,2.08) -- (0.6823,2.521) -- (0.4048,3.056) -- (0.245,3.705) -- (0.1514,4.491) -- (0.09586,5.444) -- (0.06233,6.599) -- (0.04173,8) -- (0.04173,8) -- (15,8) -- (15,0.9632) -- cycle;
\fill[cNS!45] (0.04173,8) -- (0.04173,8) -- (0.03418,8.857) -- (0.02826,9.807) -- (0.02358,10.86) -- (0.01986,12.02) -- (0.01689,13.31) -- (0.01451,14.74) -- (0.01258,16.32) -- (0.01101,18.06) -- (0.009733,20) -- (0.009733,20) -- (15,20) -- (15,8) -- cycle;
\fill[cBH!45] (0.009733,20) -- (0.009733,20) -- (0.008284,23.15) -- (0.007183,26.8) -- (0.006339,31.02) -- (0.005687,35.91) -- (0.00518,41.57) -- (0.004784,48.12) -- (0.004473,55.7) -- (0.004228,64.47) -- (0.004033,74.63) -- (0.003879,86.39) -- (0.003757,100) -- (15,100) -- (15,20) -- cycle;
\draw[white, line width=3.2pt] (15.9,0.9357) -- (8.597,1.099) -- (4.86,1.291) -- (2.868,1.517) -- (1.747,1.782) -- (1.09,2.094) -- (0.6914,2.46) -- (0.4451,2.89) -- (0.2905,3.395) -- (0.1923,3.988) -- (0.1293,4.685) -- (0.08846,5.504) -- (0.06166,6.466) -- (0.04386,7.597) -- (0.03189,8.924) -- (0.02372,10.48) -- (0.01807,12.32) -- (0.0141,14.47) -- (0.01127,17) -- (0.009227,19.97) -- (0.007734,23.46) -- (0.006628,27.56) -- (0.005802,32.38) -- (0.005179,38.04) -- (0.004706,44.69) -- (0.004344,52.5) -- (0.004066,61.68) -- (0.003851,72.46) -- (0.003685,85.12) -- (0.003556,100);
\draw[cGI, line width=2.2pt] (15.9,0.9357) -- (8.597,1.099) -- (4.86,1.291) -- (2.868,1.517) -- (1.747,1.782) -- (1.09,2.094) -- (0.6914,2.46) -- (0.4451,2.89) -- (0.2905,3.395) -- (0.1923,3.988) -- (0.1293,4.685) -- (0.08846,5.504) -- (0.06166,6.466) -- (0.04386,7.597) -- (0.03189,8.924) -- (0.02372,10.48) -- (0.01807,12.32) -- (0.0141,14.47) -- (0.01127,17) -- (0.009227,19.97) -- (0.007734,23.46) -- (0.006628,27.56) -- (0.005802,32.38) -- (0.005179,38.04) -- (0.004706,44.69) -- (0.004344,52.5) -- (0.004066,61.68) -- (0.003851,72.46) -- (0.003685,85.12) -- (0.003556,100);
\draw[white, line width=0.6pt] (1e-3,8) -- (15,8);
\draw[white, line width=0.6pt] (1e-3,20) -- (15,20);
\draw[black!60, dashed, line width=0.3pt] (1e-3,8) -- (15,8);
\draw[black!60, dashed, line width=0.3pt] (1e-3,20) -- (15,20);
\fill[white, opacity=0.45] (11,0.08) rectangle (15,100);
\fill[pattern=north east lines, pattern color=black!35] (11,0.08) rectangle (15,100);
\draw[black!85, dashed, line width=0.5pt] (10,0.08) -- (10,100);
\draw[black!85, dotted, line width=0.6pt] (11,0.08) -- (11,100);
\node[font=\scriptsize, align=center] at (0.05,1.5) {main\\sequence};
\node[font=\scriptsize, align=center] at (3.6,2.6) {white\\dwarf};
\node[font=\scriptsize, align=center] at (2.2,12.5) {neutron\\star};
\node[font=\scriptsize] at (0.55,30) {black hole};
\node[font=\scriptsize, align=center, inner sep=1pt] (agb) at (0.55,62) {giant + AGB\\($\sim12\%$ of $\tau_{\rm MS}$, not followed)};
\draw[black!80, line width=0.3pt] (agb.west) -- (0.005183,38);
\end{axis}
\end{tikzpicture}
\caption{Evolutionary stage of a star as a function of its initial mass $M_i$ and of the time $t$ since its formation, for $Z=Z_\odot$: main sequence and WD, NS and BH remnants. The thick line is the giant and AGB stage, which our model skips. The vertical lines mark the ages of the thin disc ($10\Gyr$) and of the thick disc and bulge ($11\Gyr$).}
\label{fig:phases}
\end{figure}
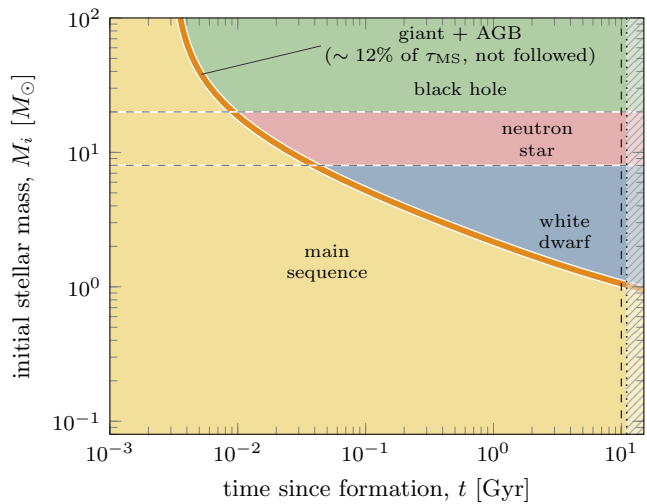

In Fig.~\ref{fig:phases} we show the stage of a star as a function of its initial mass and of the time since its formation. The two-phase treatment skips the short giant and AGB stages (thick line) and starts each remnant with an empty atom. Including those stages, and letting the remnant inherit the atom of its progenitor, changes the maximum by less than $0.1\%$, mostly because condition (i) fails for giants. We also treat all stars as single; a companion would either act as part of one heavier object or perturb the atom, effects of order unity.

The integral \eqref{eq:eps} is evaluated on a grid of 60 initial masses, eight formation episodes, two evolutionary phases and 1210 spatial nodes.

\begin{figure*}[t]
\centering
\includegraphics[width=0.95\textwidth]{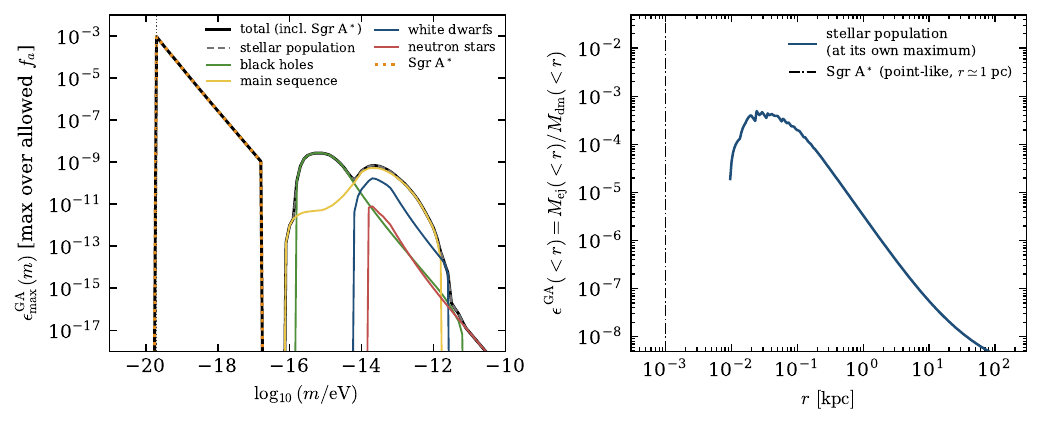}
\caption{Left: bound on $\epsGA$ maximised over the allowed values of $f_a$, split by the type of object. Right: fraction of the DM within a radius $r$ that the stellar population radiates, $\epsGA(<r)=M_{\rm ej}(<r)/M_{\rm dm}(<r)$, at its maximum; the dash-dotted line marks the radius at which the Sgr~A$^*$ term is evaluated.}
\label{fig:breakdown}
\end{figure*}

\subsection{Milky Way}
\label{sec:results}

\begin{figure}[t]
\centering
\includegraphics[width=\columnwidth]{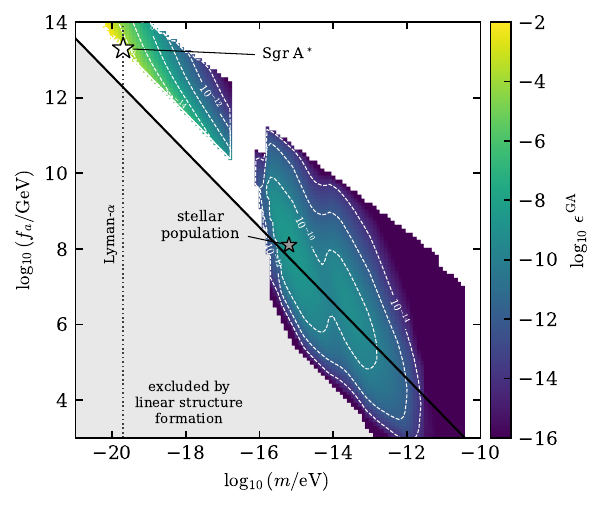}
\caption{Upper bound on the fraction $\epsGA$ of the Milky-Way DM ejected by bosenovae in $10\Gyr$, Eq.~\eqref{eq:eps}; blank regions have $\epsGA=0$, and the white dashed lines are contours of constant $\epsGA$. The grey region is excluded by Eq.~\eqref{eq:cosmo}, and the dotted line is the Lyman-$\alpha$ bound. The small and large stars mark the maxima of the stellar population and of Sgr~A$^*$.}
\label{fig:map}
\end{figure}

In Fig.~\ref{fig:map} we show the bound on $\epsGA$ over the $(m,f_a)$ plane. The stars and Sgr~A$^*$ contribute in separate windows of boson mass: the stars above $m\simeq7\times10^{-17}\eV$ (appreciably only above $1.5\times10^{-16}\eV$, where the heaviest BHs become dense), and Sgr~A$^*$ below $m\simeq2\times10^{-17}\eV$, where its $\alpha$ reaches $1/2$. For the stars, the largest value in the allowed region is
\begin{equation}
\epsGA_\star\simeq2.8\times10^{-9}\ \text{at}\ m\simeq7\times10^{-16}\eV,\, f_a\simeq1.3\times10^{8}\GeV,
\end{equation}
about $3\times10^{3}\Msun$ of DM over the age of the Galaxy. Sgr~A$^*$ raises the maximum to $\epsGA\simeq10^{-3}$ at the Lyman-$\alpha$ bound, but through the less controlled estimate of Sec.~\ref{sec:sgra}.

The shape of the allowed region follows from Sec.~\ref{sec:GAs}. Since $\GGA\McGA\propto m^{-5}f_a^{-2}$, the lines of constant $\epsGA$ follow $f_a\propto m^{-5/2}$. The sharp lower edge is condition (iii), and the left edge is the dense condition, reached by the heaviest BHs in the inner bulge. The maximum sits where these constraints meet.

\begin{table*}[t]
\caption{Maximal ejection by stars for the galaxy templates of Sec.~\ref{sec:galaxies}, over the region allowed by Eq.~\eqref{eq:cosmo}. $M_{\rm dm}$ is the halo mass, $M_{\star}$ the formed stellar mass, $\sigma_0$ the velocity added in quadrature to the circular speed, and $M_{\rm dm}(<r_{1/2})$ the DM mass within the stellar half-mass radius. The grid is coarser than that of Fig.~\ref{fig:map}, and Sgr~A$^*$ is not included.}
\label{tab:galaxies}
\footnotesize
\begin{ruledtabular}
\begin{tabular}{lccccccc}
Galaxy & $M_{\rm dm}$ & $M_\star$ & $\sigma_0$ & $\epsGA_{\max}$ & $M_{\rm ej}^{\rm tot}$ & $M_{\rm ej}^{\rm tot}/M_{\rm dm}(<r_{1/2})$ & $(\log m,\log f_a)$ \\
 & [$\Msun$] & [$\Msun$] & [km/s] & & [$\Msun$] & & [eV, GeV] \\
\hline
Milky Way (fiducial, NFW) & $1.2\times10^{12}$ & $8\times10^{10}$ & 100 & $2.7\times10^{-9}$ & $3.3\times10^{3}$ & $2.1\times10^{-7}$ & $(-15.2,\ 8.2)$ \\
MW, $r^{-1.5}$ cusp + NSC & $1.5\times10^{12}$ & $8\times10^{10}$ & 100 & $1.6\times10^{-7}$ & $2.5\times10^{5}$ & $5.2\times10^{-6}$ & $(-15.2,\ 10.2)$ \\
Fornax-like dSph (Burkert, $Z=0.1Z_\odot$) & $1.7\times10^{8}$ & $4.3\times10^{7}$ & 11.7 & $4.5\times10^{-6}$ & $7.8\times10^{2}$ & $1.2\times10^{-5}$ & $(-16.2,\ 9.4)$ \\
DF2-like UDG (NFW, $Z=0.1Z_\odot$) & $5.0\times10^{10}$ & $3\times10^{8}$ & 8.4 & $3.6\times10^{-8}$ & $1.8\times10^{3}$ & $1.0\times10^{-6}$ & $(-16.2,\ 9.8)$ \\
NGC 1277-like cE (NFW) & $2.0\times10^{13}$ & $1.8\times10^{11}$ & 317 & $6.7\times10^{-11}$ & $1.3\times10^{3}$ & $4.4\times10^{-7}$ & $(-14.8,\ 7.4)$ \\
Giant elliptical (NFW) & $5.0\times10^{13}$ & $4\times10^{11}$ & 250 & $1.8\times10^{-11}$ & $9.2\times10^{2}$ & $5.4\times10^{-9}$ & $(-15.0,\ 7.6)$ \\
\end{tabular}
\end{ruledtabular}
\end{table*}

\begin{figure*}[t]
\centering
\includegraphics[width=0.95\textwidth]{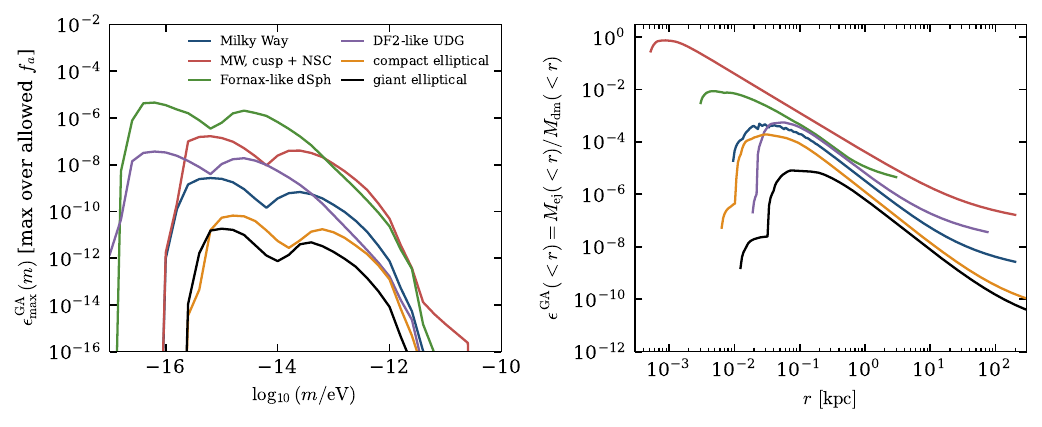}
\caption{Left: maximal ejected fraction over the allowed values of $f_a$, as a function of $m$, for the galaxy templates of Table~\ref{tab:galaxies}. Right: fraction of the DM within a radius $r$ that is radiated, $\epsGA(<r)=M_{\rm ej}(<r)/M_{\rm dm}(<r)$, at the maximum of each template. Each curve starts where the validity conditions or the stellar distribution first allow ejection, and ends where its halo is truncated.}
\label{fig:galaxies}
\end{figure*}

In Fig.~\ref{fig:breakdown} we split the result by type of object (left) and show the fraction of the DM within each radius that is radiated (right). BHs dominate at the smallest masses and give the maximum, main-sequence stars dominate between $10^{-14}$ and $10^{-12}\eV$, and WDs and NSs are subdominant. Sgr~A$^*$ dominates below $\sim10^{-17}\eV$, where its contribution falls steeply with $m$ until condition (ii) fails. Since the throughput scales as $\rhodm^2/\vdm^2$, $90\%$ of the ejected mass comes from BHs in the inner few hundred parsecs, while the solar neighbourhood, where the estimates of Ref.~\cite{Budker:2023sex} were made, contributes only $\sim10^{-5}$ of the total. Even within the central $25\pc$, where this fraction peaks, the stars radiate only $5\times10^{-4}$ of the DM. The result is therefore sensitive mostly to the poorly known DM density and velocity of the inner bulge.

At the maximum $\tcGA\ll T$ and the stimulated term dominates. The flux cap is inactive there: it acts only for $m\gtrsim10^{-15}\eV$ at the smallest allowed $f_a$, where it lowers the maximum over $f_a$ by at most $15\%$.

\subsubsection{The supermassive black hole}
\label{sec:sgra}

One object of the Galaxy deserves separate treatment: Sgr~A$^*$, with $M_{\rm BH}^{\rm Sgr\,A^*}=4\times10^{6}\Msun$ \cite{GRAVITY:2018ofz}. Its atom is dense down to $m\simeq2\times10^{-21}\eV$, has a size of order a parsec, and sits in the densest DM of the Galaxy, $\rhodm\gtrsim10^{3}\,{\rm GeV/cm^3}$ within the central parsec for an NFW cusp. For the masses allowed by the Lyman-$\alpha$ forest and by Eq.~\eqref{eq:cosmo}, the throughput at the smallest allowed $f_a$ is limited by the geometric flux, dominated by gravitational focusing, rather than by the perturbative rate. The bound is then
\begin{equation}
M_{\rm ej}(T)\lesssim\Fgeo T\simeq\frac{2\pi\rhodm T}{m^{2}\vdm},
\label{eq:Mej_sgra}
\end{equation}
which does not depend on the mass of the BH. For an NFW cusp at $1\pc$ and $m=2\times10^{-20}\eV$ this is $\simeq10^{9}\Msun$ in $10\Gyr$, reached for $f_a\simeq2\times10^{13}\GeV$: a fraction $\epsGA\simeq10^{-3}$ of the halo, comparable to all the DM within the inner kiloparsec.

This is only a geometric upper bound. At these parameters the perturbative rate exceeds the flux, so the treatment of Ref.~\cite{Budker:2023sex} has broken down (Sec.~\ref{sec:GAs}); the flux assumes that the DM around the BH is continuously renewed; and the density and velocity of that DM are themselves uncertain. Superradiance plays no role, since $\alpha\simeq6\times10^{-4}$ at the maximum. Moreover, $m\simeq2\times10^{-20}\eV$ is disfavoured by the stellar kinematics of ultrafaint dwarf galaxies, which require $m\gtrsim3\times10^{-19}\eV$ \cite{Dalal:2022rmp}, and by stellar orbits around Sgr~A$^*$ \cite{Bar:2019pnz}. Above $3\times10^{-19}\eV$ the Sgr~A$^*$ term is at most $3\times10^{-6}$. We therefore regard it as an illustration of the potential relevance of supermassive BHs rather than as a robust bound.

\subsection{Other galaxies}
\label{sec:other}
\label{sec:galaxies}

The Milky Way is not necessarily the galaxy in which the pump is most efficient, and the question of the Introduction concerns in the first place galaxies that appear to lack DM. For a given object the throughput is largest at the edge of the validity region, $\xifoc\simeq1$ and the smallest allowed $f_a$, where it grows steeply with the mass of the object, with the DM density and with the inverse of the DM velocity. The environments to examine are therefore low-velocity systems, such as dwarf spheroidals and ultra-diffuse galaxies, massive BHs and dense nuclei.

We have repeated the calculation for the templates of Table~\ref{tab:galaxies}. Their stars are Hernquist spheroids formed in a single burst, with masses, sizes, ages, metallicities and velocity dispersions taken from observations. The Fornax-like dwarf spheroidal has the cored Burkert halo \cite{Burkert:1995yz} fitted to Fornax, $\rho_0=0.19\Msun/{\rm pc}^3$ and $r_0=0.37\kpc$ \cite{Salucci:2011ee}, and the stellar mass \cite{deBoer:2012py}, size, metallicity and velocity dispersion \cite{McConnachie:2012vd} of Fornax; we take its stars to be $10\Gyr$ old, older than most of them \cite{deBoer:2012py}, which lengthens the life of their remnants. The DF2-like ultra-diffuse galaxy has the stars of NGC~1052-DF2 \cite{vanDokkum:2018vup,Fensch:2019,Danieli:2019zyi}, and the NGC~1277-like compact elliptical those of NGC~1277 \cite{Trujillo:2013ana,Yildirim:2015zxa}. Both sit in the NFW halos they would have had before any depletion, from abundance matching \cite{Behroozi:2012iw,vanDokkum:2018vup}, with the concentrations of Ref.~\cite{Dutton:2014xda}. The giant elliptical follows the scaling relations of massive early-type galaxies for its halo \cite{Kravtsov:2014sra,Dutton:2014xda}, size \cite{Shen:2003sda}, velocity dispersion \cite{Zahid:2016} and stellar populations \cite{Thomas:2004ep}. Finally, a more favourable Milky Way has a steeper inner cusp, $\rho\propto r^{-1.5}$ \cite{Moore:1999gc}, and a nuclear star cluster of $3\times10^{7}\Msun$ with a half-light radius of $4.2\pc$ \cite{Schodel:2014gna} and old stars \cite{Schodel:2020}.

In Fig.~\ref{fig:galaxies} we show the results, from which three conclusions follow. First, every template ejects about $10^{3}\Msun$: dwarfs have a larger yield per star, because of their low velocities, but far fewer stars. The largest fraction of a halo, $5\times10^{-6}$, is reached in the dwarf spheroidal, where the first effect wins. Second, the pump is locally significant only in a dense nucleus (Fig.~\ref{fig:galaxies}, right): the BHs of a nuclear star cluster in a steep cusp can process $\sim10\%$ of the DM within the central $5\pc$ in $10\Gyr$, and about three quarters of it within the central parsec, while in the other templates the fraction stays below $10^{-2}$ at every radius. Since this DM is replenished on a crossing time, the cusp is not emptied but steadily converted into radiation. Third, no template comes within four orders of magnitude of the depletion needed to explain a DM-deficient galaxy: the DF2-like one ejects $\sim10^{3}\Msun$, against the $\sim5\times10^{10}\Msun$ halo it would have to have lost.

\subsubsection{Central black holes of other galaxies}
\label{sec:otherbh}

\begin{figure*}[t]
\centering
\includegraphics[width=0.95\textwidth]{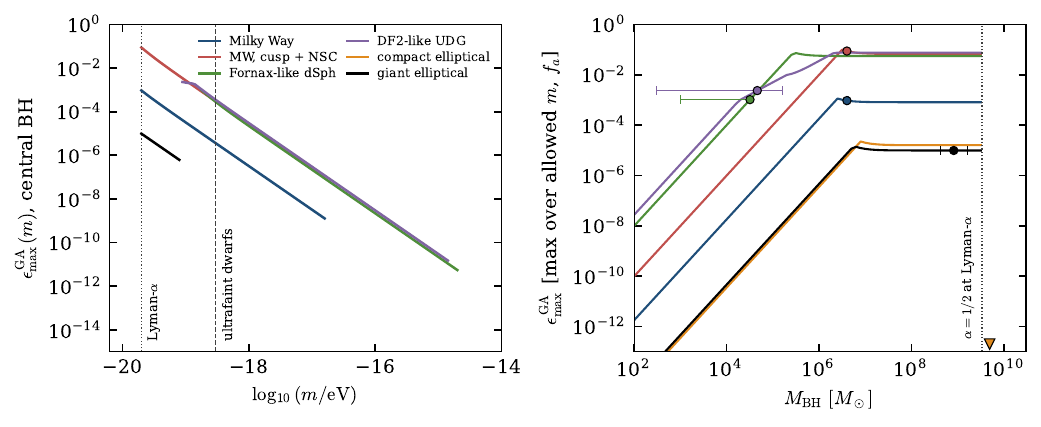}
\caption{Bound on the fraction of the halo radiated by the atom of the central BH of each template of Table~\ref{tab:galaxies}. Left: maximum over the allowed $f_a$, as a function of $m$, for the BH masses of Sec.~\ref{sec:otherbh}; the compact elliptical gives nothing. The dotted and dashed lines are the Lyman-$\alpha$ bound and the bound from ultrafaint dwarfs. Right: maximum over the allowed $m$ and $f_a$, as a function of the BH mass; the points mark the adopted masses and the bars their range, and the dotted line the mass above which $\alpha>1/2$ at the Lyman-$\alpha$ bound.}
\label{fig:centralbh}
\end{figure*}

The estimate of Sec.~\ref{sec:sgra} applies to the central BH of any galaxy. We give each template of Table~\ref{tab:galaxies} a BH of realistic mass: Sgr~A$^*$ for both Milky Ways; $3.2\times10^{4}\Msun$ for the dwarf spheroidal, the upper limit in Fornax \cite{Jardel:2012}; $5\times10^{4}\Msun$ for the DF2-like galaxy, which has no measured BH, from the relation between BH and stellar mass of Ref.~\cite{Reines:2015} (from $3\times10^{2}$ to $2\times10^{5}\Msun$ within its scatter and the extrapolated $M_{\rm BH}$--$\sigma$ relation); $4.9\times10^{9}\Msun$ for the compact elliptical, as measured in NGC~1277 \cite{Walsh:2016} ($1.2\times10^{9}\Msun$ in Ref.~\cite{Graham:2016}); and $8\times10^{8}\Msun$ for the giant elliptical, from the $M_{\rm BH}$--$\sigma$ relation \cite{Kormendy:2013dxa}. We take the density and velocity of the DM of each halo at the larger of $R_*$ and $1\pc$, and require the atom to lie within the sphere of influence of the BH, $R_*<GM_{\rm BH}/\sigma^2$, where its potential is that of a point mass. For Sgr~A$^*$ this reproduces the estimate of Sec.~\ref{sec:sgra}.

Figure~\ref{fig:centralbh} (right) shows three regimes. Light BHs have small spheres of influence, so their atoms fit only at $m\gtrsim\sigma/M_{\rm BH}$, and since the flux falls as $m^{-2}$, $\epsGA$ grows as $M_{\rm BH}^2$. Between $\sim10^{6}$ and $3\times10^{9}\Msun$ the atom fits at the Lyman-$\alpha$ bound and the throughput is the geometric flux of Eq.~\eqref{eq:Mej_sgra}, independent of $M_{\rm BH}$: the ejected mass depends only on $\rhodm/\vdm$ near the centre, and the fraction is largest in dense and cold haloes, $5$--$8\%$ in the dwarfs and in the Milky Way with a cusp. Above $3\times10^{9}\Msun$, $\alpha>1/2$ at the Lyman-$\alpha$ bound and no atom forms at allowed masses, so the BH of NGC~1277 contributes nothing. With the adopted masses, the BHs of the dwarf spheroidal and of the DF2-like galaxy radiate $1\times10^{-3}$ and $2\times10^{-3}$ of their haloes, at $m\simeq2\times10^{-19}$ and $9\times10^{-20}\eV$, comparable to Sgr~A$^*$ in the Milky Way; the giant elliptical radiates $10^{-5}$, at $\alpha\simeq0.1$, where superradiance of a spinning BH could compete; and Sgr~A$^*$ in the steeper cusp $0.09$. Since the BHs of the two dwarfs lie on the rising part of the curve, the uncertainty in their masses dominates the result: over the ranges marked by the bars, $\epsGA$ spans $10^{-6}$--$10^{-3}$ for the dwarf spheroidal and $2\times10^{-7}$--$8\times10^{-3}$ for the DF2-like galaxy, whereas for the Milky Way and the giant elliptical, whose BHs lie on the plateau, it changes by less than $3\%$. Above $3\times10^{-19}\eV$ the dwarfs give at most $3\times10^{-4}$.

These numbers share the caveats of Sec.~\ref{sec:sgra}, in a more extreme form: the BH would have to radiate all the DM within $60\pc$ of the dwarf spheroidal, $0.6\kpc$ of the DF2-like galaxy and $10\kpc$ of the Milky Way with a cusp, which the flux assumes to be continuously renewed. Even so, a DF2-like galaxy with a realistic BH would lose at most $\sim4\times10^{8}\Msun$, two orders of magnitude below the $5\times10^{10}\Msun$ it would have to lose.

\section{Boson star pumps}
\label{sec:bosestars}

\begin{figure}[t]
\centering
\includegraphics[width=\columnwidth]{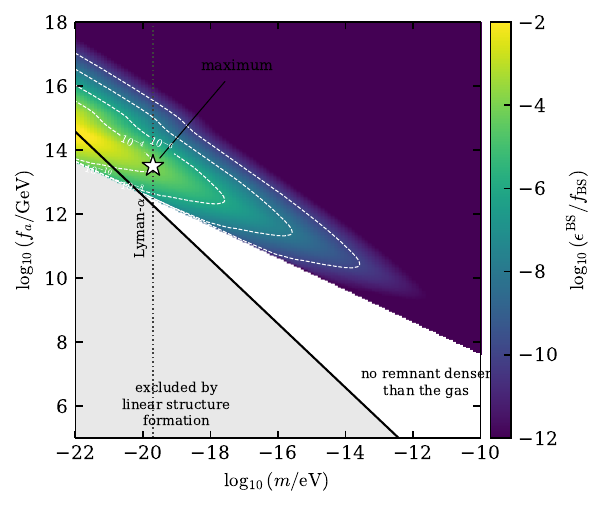}
\caption{Fraction $\epsBS/\fBS$ of the Milky-Way DM radiated by BSs per unit $\fBS$ in $10\Gyr$, Eq.~\eqref{eq:epsBS}, for $\fBS\ll1$, where the gas is the whole halo; the white dashed lines are contours of constant $\epsBS/\fBS$. In the white region no remnant is denser than the gas, condition (a). The grey region is excluded by Eq.~\eqref{eq:cosmo} and the dotted line is the Lyman-$\alpha$ bound. The star marks the maximum.}
\label{fig:bsmap}
\end{figure}

\begin{figure*}[t]
\centering
\includegraphics[width=0.95\textwidth]{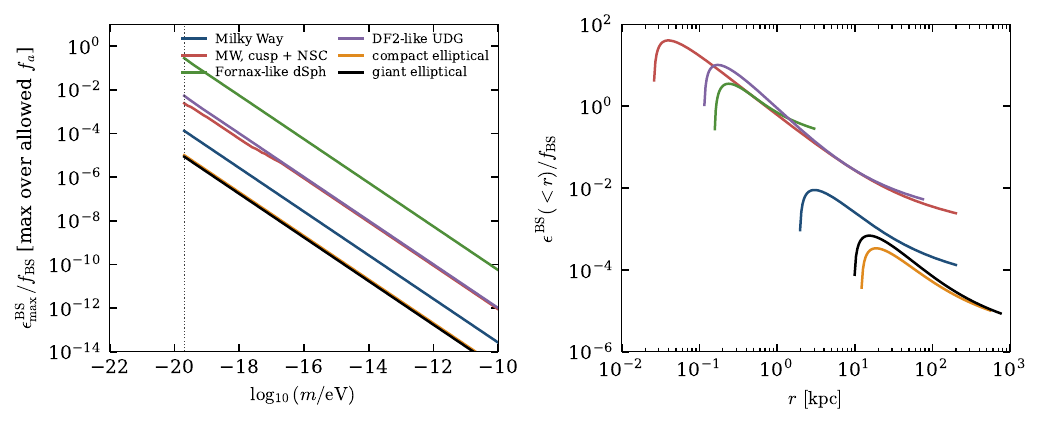}
\caption{Left: maximum of $\epsBS/\fBS$ over the allowed decay constants, as a function of $m$, for $\fBS\ll1$ and the templates of Sec.~\ref{sec:galaxies}; the dotted line is the Lyman-$\alpha$ bound. Right: fraction of the DM within a radius $r$ that is radiated, per unit $\fBS$, $\epsBS(<r)/\fBS$, at the maxima of the left panel.}
\label{fig:bsbreak}
\end{figure*}

We now assume that a fraction $\fBS$ of the DM of a galaxy is in BSs, which may have formed in the early Universe or by relaxation of the halo (Sec.~\ref{sec:BSmodel}). The rest of the DM, with density $(1-\fBS)\rhodm$ and the velocity of the halo, is the gas they feed on, and we neglect its depletion within $T$. Each star cycles as described in Sec.~\ref{sec:BSmodel}. A star that starts with a mass $M_0^{\rm BS}$ absorbs $\McBS-M_0^{\rm BS}$ before its first bosenova, at $\tcBS$, and afterwards never falls below the remnant mass $(1-\dBS)\McBS$. In a time $T>\tcBS$ it therefore radiates at most
\begin{equation}
\MejBS=\underbrace{\dBS\McBS}_{\rm store}+\underbrace{(T-\tcBS)\min\{\GBS\McBS,\Fgeo\}}_{\rm pump},
\label{eq:MejBS}
\end{equation}
and nothing if $\tcBS>T$, with $\Fgeo$ the flux through the gravitationally focused cross section of the star. As for an atom:
\begin{itemize}
\item \emph{Store}: the mass that the star holds above its remnant, which it can radiate at most once; it replaces the $\McGA$ of Eq.~\eqref{eq:bound} because, unlike an atom, a star keeps a remnant.
\item \emph{Pump}: the DM that the star takes from the gas after $\tcBS$ and radiates, at the throughput $\GBS\McBS$ capped by $\Fgeo$.
\end{itemize}
We treat them separately, because they depend on different unknowns.

Besides Eqs.~\eqref{eq:stab} and~\eqref{eq:cosmo} and the Lyman-$\alpha$ bound, we impose three conditions specific to the stars:
\begin{enumerate}
\item[(a)] \emph{Distinct stars}: the stars are denser than the gas along their whole cycle, i.e.\ already as remnants, which requires $f_a^4\gtrsim\rhodm/(8m^2)$ (Sec.~\ref{sec:BSmodel}).
\item[(b)] \emph{Room}: the galactocentric radius exceeds both the de Broglie wavelength of the gas and the size of the remnant.
\item[(c)] \emph{Growth}: $\GBS>0$, so that the remnant grows back to $\McBS$.
\end{enumerate}

The pump depends on the initial mass through $\tcBS$, Eq.~\eqref{eq:tcBS}, and this mass is set by the formation of the stars, which is not known. We take a stationary population, in which the masses of the stars are spread along their cycle and their total mass does not change. The stars then radiate exactly the mass they absorb, at most $\GBS\McBS$ per unit time each, and the pump term holds with $\tcBS=0$. We have seen that the cap is never active for BSs, since the stars are large and dilute, and $\GBS\McBS/\Fgeo\lesssim3\times10^{-2}$ at all the maxima below. The stars follow the DM, with $\fBS\rhodm(\bm x)/\McBS$ of them per unit volume, so that $\fBS$ is the fraction of the DM that they would hold at $\McBS$. Summing the pump term over them, as in Eq.~\eqref{eq:eps}, the halo loses a fraction
\begin{equation}
\begin{split}
\epsBS\equiv\frac{M_{\rm ej}^{\rm tot}}{M_{\rm dm}}&=\frac{1}{M_{\rm dm}}\int d^{3}x\,\frac{\fBS\rhodm(\bm x)}{\McBS}\\
&\quad\times\MejBS\big(T;(1-\fBS)\rhodm(\bm x),\vdm(\bm x)\big),
\end{split}
\label{eq:epsBS}
\end{equation}
where $\MejBS$ is the pump term of Eq.~\eqref{eq:MejBS} with $\tcBS=0$, for a star in the gas density $(1-\fBS)\rhodm(\bm x)$ and velocity $\vdm(\bm x)$, and vanishes wherever Eq.~\eqref{eq:stab} or conditions (a)--(c) fail. Per unit mass in stars the integrand is the local efficiency $\epsBS_{\rm loc}=\GBS T$, the fraction of its own mass that a star pumps in $T$, which can exceed unity because the gas is never exhausted. The store term of Eq.~\eqref{eq:MejBS} would add at most $\dBS\fBS$, if all the stars happened to be at $\McBS$ when the time $T$ starts; we discuss it with the results. We first take $\fBS\ll1$, where the gas is the whole halo and $\epsBS/\fBS$ does not depend on $\fBS$, and then larger fractions.

\subsection{Results}
\label{sec:bsresults}

In Fig.~\ref{fig:bsmap} we show $\epsBS/\fBS$ in the Milky Way. At every radius we compute $\epsBS_{\rm loc}=\GBS T$, with the full rate $\GBS=\Gamma_{\rm gr}+\Gamma_\lambda$ of Eq.~\eqref{eq:rate} at $\McBS$, wherever conditions (a)--(c) and Eq.~\eqref{eq:stab} hold. Gravity makes the heaviest stars grow only at the largest $f_a$ (Sec.~\ref{sec:BSmodel}), where it is inefficient, and makes light stars evaporate at smaller $f_a$. Self-interactions dominate at smaller $f_a$, down to the edge of the white region, below which condition (a) fails. The efficiency is largest just above this edge, where $\Gamma_\lambda\propto\rhodm/(m\vdm^2)$, and at the smallest $m$: $\epsBS/\fBS\simeq10^{-4}$ at the Lyman-$\alpha$ bound and $f_a\simeq3\times10^{13}\GeV$.

In Fig.~\ref{fig:bsbreak} (left) we compare the galaxy templates of Sec.~\ref{sec:galaxies}. Denser and colder systems do better: $\epsBS/\fBS\simeq2\times10^{-3}$ for the Milky Way with a cusp, $0.5\%$ for the DF2-like galaxy and $28\%$ for the dwarf spheroidal, all at the Lyman-$\alpha$ bound. For heavier bosons the maximum follows the same edge and falls as $1/m$.

In Fig.~\ref{fig:bsbreak} (right) we show the fraction of the DM within each radius that is radiated, per unit $\fBS$, at these maxima. It peaks just outside the inner edge of the region where the model applies: $\sim10^{-2}$ within $3\kpc$ in the Milky Way, $3.5$ within $0.2\kpc$ in the dwarf, $10$ within $0.2\kpc$ in the DF2-like galaxy, $40$ within $40\pc$ in the cusp and $\lesssim10^{-3}$ within $\sim20\kpc$ in the ellipticals, whose inner $10\kpc$ fail condition (a). There the local efficiency exceeds unity only in the dwarf.

\begin{figure*}[t]
\centering
\includegraphics[width=0.97\textwidth]{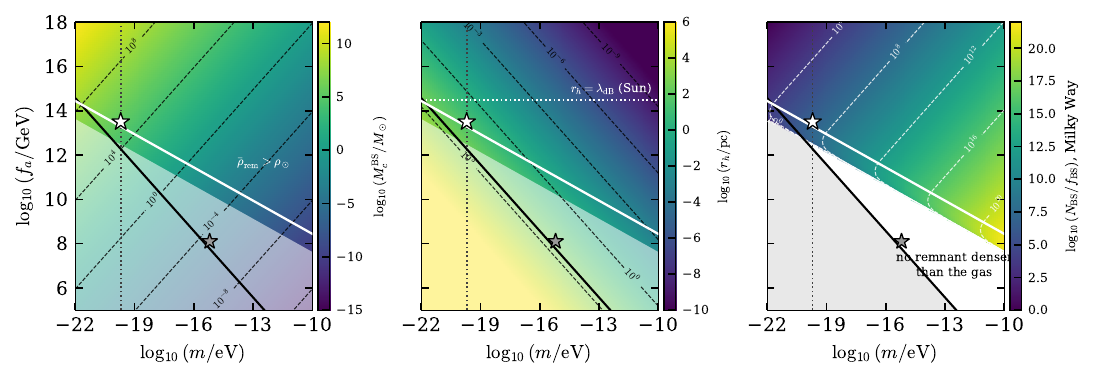}
\caption{Properties of the BSs at their maximum mass: mass $\McBS$ (left), half-mass radius $r_h$ (centre), and number in the Milky Way per unit $\fBS$, counting only the regions where their remnants are denser than the gas (right). There are none in the white region, which is faded in the other panels. The white line marks where the remnants are as dense as the DM at the Sun, the black line is the cosmological bound and the dotted line the Lyman-$\alpha$ bound; in the centre panel, a second dotted line marks where $r_h$ equals the de Broglie wavelength at the Sun. The white and grey stars mark the maxima of Figs.~\ref{fig:bsmap} and~\ref{fig:map}.}
\label{fig:bsprops}
\end{figure*}

In Fig.~\ref{fig:bsprops} we show the mass, size and number of the stars. Stars whose remnants are denser than the gas exist only above the line $f_a^4\simeq\rhodm/(8m^2)$ (white line), far above the decay constants at which stellar bosenovae are most efficient (grey star). At the maxima of Fig.~\ref{fig:bsbreak} (white star for the Milky Way) the stars are heavy and extended, $\McBS\sim10^{5}$--$10^{6}\Msun$ and $r_h\sim3$--$30\pc$, and the Milky Way would hold $\sim10^{7}\fBS$ of them. Their remnants are 2.4 times wider, and near the white line, where they are barely denser than the gas, they overlap unless $\fBS\lesssim0.1$.

For larger $\fBS$, more stars leave less gas. At the optimal $f_a$, $\epsBS_{\rm loc}$ is proportional to the density of the gas, so $\epsBS\simeq \fBS(1-\fBS)$ times the values found above for $\epsBS/\fBS$. It is largest at $\fBS\simeq1/2$, where it is about a quarter of them: $3\times10^{-5}$ for the Milky Way, $6\times10^{-4}$ with a cusp, $0.1\%$ for the DF2-like galaxy and $6\%$ for the dwarf spheroidal. Requiring the remnants not to overlap, $\fBS\lesssim0.1$, lowers these values by another factor of three.

\begin{figure*}[!t]
\centering
\includegraphics[width=0.95\textwidth]{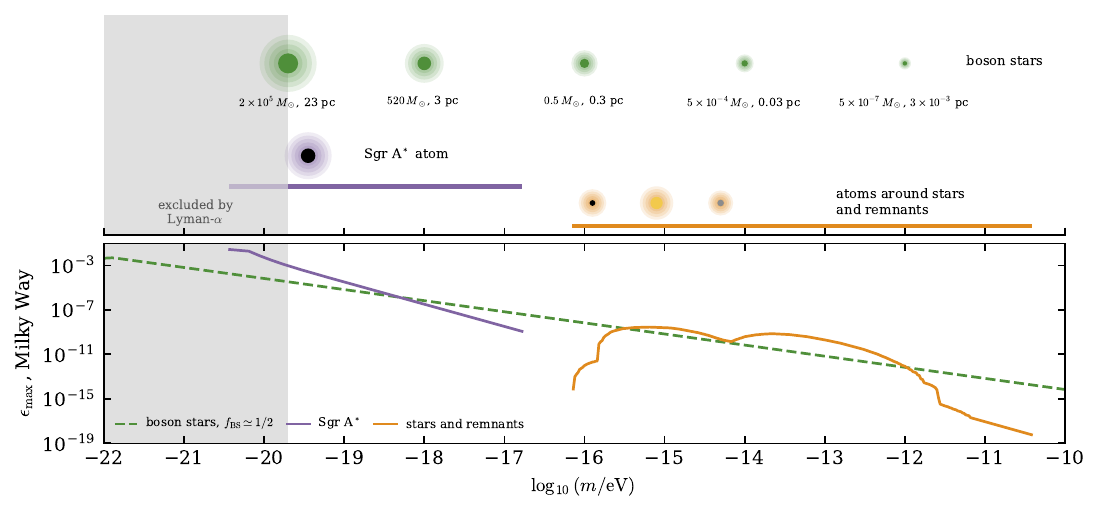}
\caption{Overview of the channels as a function of the boson mass. Top: where each channel acts: BSs (drawn with the mass and size they have where they pump best), the atom of Sgr~A$^*$, and the atoms of stars and remnants; the bars span the masses at which each kind of atom ejects DM, i.e.\ the full extent of the curves below. Bottom: largest fraction of the Milky-Way halo radiated in $10\Gyr$ by each channel, over the decay constants allowed by the cosmological bound; for BSs, also over $\fBS$, which gives $\fBS\simeq1/2$. Masses below $2\times10^{-20}\eV$ (grey) are excluded by the Lyman-$\alpha$ forest.}
\label{fig:overview}
\end{figure*}

Even with the most favourable fraction of stars and a gas that is never depleted, the pump of a Milky-Way-like halo would therefore radiate at most $3\times10^{-5}$ of its mass, a DF2-like galaxy $0.1\%$ and a dwarf spheroidal $6\%$, all for $m$ close to the Lyman-$\alpha$ bound and half of the DM in BSs. 

The store term of Eq.~\eqref{eq:MejBS} is of a different nature. Stars that are within a fraction $\epsBS_{\rm loc}$ of their maximum mass when the time $T$ starts collapse once and radiate up to a fraction $\dBS$ of their own mass, whatever the gas. If all of them were there, the halo would lose $\dBS\fBS\simeq0.15$ of its mass for $\fBS=1/2$; in a stationary population only a fraction $\epsBS_{\rm loc}/\ln[1/(1-\dBS)]$ of them are, and their collapse is already part of the pump. This contribution is therefore bounded by $\dBS\fBS$, and it is set by the formation of the stars rather than by their feeding on the halo.

The main uncertainty is $\Gamma_\lambda$: the stars that dominate have $k_gr_s\sim2$--$30$ at $\McBS$ and up to $\sim70$ as remnants, beyond the range in which Eq.~\eqref{eq:Glam} was derived. By the argument of Sec.~\ref{sec:BSmodel} their true rate is smaller, so these numbers are upper estimates; a simulation of the growth of such stars would turn them into predictions.

\section{Discussion and conclusions}
\label{sec:conclusions}

We have asked whether the bosenovae of self-interacting ultralight DM, captured by stars and compact objects into gravitational atoms or condensed into boson stars, can deplete the DM of a galaxy over cosmic time. For every channel we have examined the answer is no, by a wide margin.

For the atoms the argument is simple. An atom radiates at most what it captures, and its throughput grows steeply towards small $m$ and $f_a$. Summed over a realistic model of the Milky Way, with stellar evolution and compact remnants, this bound gives $\epsGA_\star\simeq3\times10^{-9}$ of the halo in $10\Gyr$, about $3\times10^{3}\Msun$, almost all of it from stellar-mass BHs within a few hundred parsecs of the centre and at $m\simeq7\times10^{-16}\eV$. Other galaxies, from dwarf spheroidals to giant ellipticals, eject about $10^{3}\Msun$ each. The result is insensitive to the details of the stellar model and only moderately sensitive to the masses of the BHs, but it rests on the poorly known density and velocity of the DM in the inner bulge. It is locally significant only in a dense nucleus, where the BHs of a nuclear star cluster can process a tenth of the DM within the central parsecs.

The supermassive BHs do better, in a separate window of boson mass. Near the Lyman-$\alpha$ bound the atom of Sgr~A$^*$ could radiate up to $10^{-3}$ of the halo, and the central BHs of dwarf galaxies similar fractions, but only for $m\simeq2\times10^{-20}\eV$, a mass disfavoured by the kinematics of ultrafaint dwarfs and by stellar orbits around Sgr~A$^*$, and through a geometric bound on the DM flux that applies where the perturbative treatment has broken down. Above $3\times10^{-19}\eV$ these terms stay below $3\times10^{-4}$. Boson stars feeding on the halo radiate at most $3\times10^{-5}$ of a Milky-Way halo, $0.1\%$ of a DF2-like one and $6\%$ of a dwarf spheroidal, again close to the Lyman-$\alpha$ bound and with half of the DM in stars. Stars that formed at their maximum mass would, in addition, collapse once and radiate up to $\dBS\fBS$ of the halo, a contribution set by their formation rather than by their feeding on the halo. Figure~\ref{fig:overview} collects the three channels for the Milky Way.

The pump may nevertheless be observable if the boson couples, even weakly, to the Standard Model. Each bosenova emits a burst of relativistic bosons, which reaches a detector as a wave train \cite{Levkov:2016rkk,Eby:2021ece}. For derivative couplings to spins such a burst exceeds cold DM of the same energy density, which makes magnetometer networks especially sensitive to it \cite{Eby:2021ece}. Although the flux from the Galaxy as a whole is negligible at the Earth, for $1.7\times10^{-14}\eV<m<1.9\times10^{-13}\eV$ the Sun hosts its own atom, whose explosions could be detectable \cite{Budker:2023sex}. 

A final remark concerns the origin of the DM. The decay constants at which every channel peaks, from $f_a\sim10^{8}\GeV$ for the atoms of stars and remnants to $10^{13}$--$10^{14}\GeV$ for Sgr~A$^*$ and the boson stars, lie many orders of magnitude below the values for which vacuum misalignment alone produces the observed DM abundance. Most of the region of the $(m,f_a)$ plane explored here therefore requires a different production mechanism, and on the misalignment line itself, where all the rates are smaller still, every channel is more negligible than the numbers above. 

\begin{acknowledgments}
We thank David J.~E. Marsh, Xisco Jimenez Forteza and Carlos Palenzuela Luque for discussions. S.G.G.\ acknowledges the hospitality of the Oskar Klein Centre, where part of this work was carried out during the making of his master thesis. The work of S.G.G.\ was supported by the Projects No.~PID2022-138963NB-I00 and PID2025-168215NB-I00, funded by the Spanish Ministry of Science, Innovation and Universities (MCIN/AEI/10.13039/501100011033) and by the Comunitat Aut\`onoma de les Illes Balears through an ``Ajut per a projectes de recerca cient\'ifica i tecnol\`ogica (PRD2025, PRD2025\_099)''. The work of J.E.\ was supported by the Swedish Research Council (VR) under grants 2018-03641 and 2019-02337. During the preparation of this work the authors made extensive use of Claude (Anthropic), an AI assistant. It was used to develop and debug the numerical code, to run calculations, to produce several of the figures, to draft and revise substantial parts of the text, and to check the bibliography. The authors have reviewed and verified all of the code, calculations and text, and take full responsibility for the content of the manuscript.
\end{acknowledgments}

\bibliographystyle{apsrev4-2}
\bibliography{refs}

\begin{thebibliography}{59}%
\makeatletter
\providecommand \@ifxundefined [1]{%
 \@ifx{#1\undefined}
}%
\providecommand \@ifnum [1]{%
 \ifnum #1\expandafter \@firstoftwo
 \else \expandafter \@secondoftwo
 \fi
}%
\providecommand \@ifx [1]{%
 \ifx #1\expandafter \@firstoftwo
 \else \expandafter \@secondoftwo
 \fi
}%
\providecommand \natexlab [1]{#1}%
\providecommand \enquote  [1]{``#1''}%
\providecommand \bibnamefont  [1]{#1}%
\providecommand \bibfnamefont [1]{#1}%
\providecommand \citenamefont [1]{#1}%
\providecommand \href@noop [0]{\@secondoftwo}%
\providecommand \href [0]{\begingroup \@sanitize@url \@href}%
\providecommand \@href[1]{\@@startlink{#1}\@@href}%
\providecommand \@@href[1]{\endgroup#1\@@endlink}%
\providecommand \@sanitize@url [0]{\catcode `\\12\catcode `\$12\catcode `\&12\catcode `\#12\catcode `\^12\catcode `\_12\catcode `\%12\relax}%
\providecommand \@@startlink[1]{}%
\providecommand \@@endlink[0]{}%
\providecommand \url  [0]{\begingroup\@sanitize@url \@url }%
\providecommand \@url [1]{\endgroup\@href {#1}{\urlprefix }}%
\providecommand \urlprefix  [0]{URL }%
\providecommand \Eprint [0]{\href }%
\providecommand \doibase [0]{https://doi.org/}%
\providecommand \selectlanguage [0]{\@gobble}%
\providecommand \bibinfo  [0]{\@secondoftwo}%
\providecommand \bibfield  [0]{\@secondoftwo}%
\providecommand \translation [1]{[#1]}%
\providecommand \BibitemOpen [0]{}%
\providecommand \bibitemStop [0]{}%
\providecommand \bibitemNoStop [0]{.\EOS\space}%
\providecommand \EOS [0]{\spacefactor3000\relax}%
\providecommand \BibitemShut  [1]{\csname bibitem#1\endcsname}%
\let\auto@bib@innerbib\@empty
\bibitem [{\citenamefont {Hui}\ \emph {et~al.}(2017)\citenamefont {Hui}, \citenamefont {Ostriker}, \citenamefont {Tremaine},\ and\ \citenamefont {Witten}}]{Hui:2016ltb}%
  \BibitemOpen
  \bibfield  {author} {\bibinfo {author} {\bibfnamefont {L.}~\bibnamefont {Hui}}, \bibinfo {author} {\bibfnamefont {J.~P.}\ \bibnamefont {Ostriker}}, \bibinfo {author} {\bibfnamefont {S.}~\bibnamefont {Tremaine}},\ and\ \bibinfo {author} {\bibfnamefont {E.}~\bibnamefont {Witten}},\ }\href {https://doi.org/10.1103/PhysRevD.95.043541} {\bibfield  {journal} {\bibinfo  {journal} {Phys. Rev. D}\ }\textbf {\bibinfo {volume} {95}},\ \bibinfo {pages} {043541} (\bibinfo {year} {2017})},\ \Eprint {https://arxiv.org/abs/1610.08297} {arXiv:1610.08297} \BibitemShut {NoStop}%
\bibitem [{\citenamefont {Peccei}\ and\ \citenamefont {Quinn}(1977)}]{Peccei:1977hh}%
  \BibitemOpen
  \bibfield  {author} {\bibinfo {author} {\bibfnamefont {R.~D.}\ \bibnamefont {Peccei}}\ and\ \bibinfo {author} {\bibfnamefont {H.~R.}\ \bibnamefont {Quinn}},\ }\href {https://doi.org/10.1103/PhysRevLett.38.1440} {\bibfield  {journal} {\bibinfo  {journal} {Phys. Rev. Lett.}\ }\textbf {\bibinfo {volume} {38}},\ \bibinfo {pages} {1440} (\bibinfo {year} {1977})}\BibitemShut {NoStop}%
\bibitem [{\citenamefont {Weinberg}(1978)}]{Weinberg:1977ma}%
  \BibitemOpen
  \bibfield  {author} {\bibinfo {author} {\bibfnamefont {S.}~\bibnamefont {Weinberg}},\ }\href {https://doi.org/10.1103/PhysRevLett.40.223} {\bibfield  {journal} {\bibinfo  {journal} {Phys. Rev. Lett.}\ }\textbf {\bibinfo {volume} {40}},\ \bibinfo {pages} {223} (\bibinfo {year} {1978})}\BibitemShut {NoStop}%
\bibitem [{\citenamefont {Wilczek}(1978)}]{Wilczek:1977pj}%
  \BibitemOpen
  \bibfield  {author} {\bibinfo {author} {\bibfnamefont {F.}~\bibnamefont {Wilczek}},\ }\href {https://doi.org/10.1103/PhysRevLett.40.279} {\bibfield  {journal} {\bibinfo  {journal} {Phys. Rev. Lett.}\ }\textbf {\bibinfo {volume} {40}},\ \bibinfo {pages} {279} (\bibinfo {year} {1978})}\BibitemShut {NoStop}%
\bibitem [{\citenamefont {Arvanitaki}\ \emph {et~al.}(2010)\citenamefont {Arvanitaki}, \citenamefont {Dimopoulos}, \citenamefont {Dubovsky}, \citenamefont {Kaloper},\ and\ \citenamefont {March-Russell}}]{Arvanitaki:2009fg}%
  \BibitemOpen
  \bibfield  {author} {\bibinfo {author} {\bibfnamefont {A.}~\bibnamefont {Arvanitaki}}, \bibinfo {author} {\bibfnamefont {S.}~\bibnamefont {Dimopoulos}}, \bibinfo {author} {\bibfnamefont {S.}~\bibnamefont {Dubovsky}}, \bibinfo {author} {\bibfnamefont {N.}~\bibnamefont {Kaloper}},\ and\ \bibinfo {author} {\bibfnamefont {J.}~\bibnamefont {March-Russell}},\ }\href {https://doi.org/10.1103/PhysRevD.81.123530} {\bibfield  {journal} {\bibinfo  {journal} {Phys. Rev. D}\ }\textbf {\bibinfo {volume} {81}},\ \bibinfo {pages} {123530} (\bibinfo {year} {2010})},\ \Eprint {https://arxiv.org/abs/0905.4720} {arXiv:0905.4720} \BibitemShut {NoStop}%
\bibitem [{\citenamefont {Levkov}\ \emph {et~al.}(2018)\citenamefont {Levkov}, \citenamefont {Panin},\ and\ \citenamefont {Tkachev}}]{Levkov:2018kau}%
  \BibitemOpen
  \bibfield  {author} {\bibinfo {author} {\bibfnamefont {D.~G.}\ \bibnamefont {Levkov}}, \bibinfo {author} {\bibfnamefont {A.~G.}\ \bibnamefont {Panin}},\ and\ \bibinfo {author} {\bibfnamefont {I.~I.}\ \bibnamefont {Tkachev}},\ }\href {https://doi.org/10.1103/PhysRevLett.121.151301} {\bibfield  {journal} {\bibinfo  {journal} {Phys. Rev. Lett.}\ }\textbf {\bibinfo {volume} {121}},\ \bibinfo {pages} {151301} (\bibinfo {year} {2018})},\ \Eprint {https://arxiv.org/abs/1804.05857} {arXiv:1804.05857} \BibitemShut {NoStop}%
\bibitem [{\citenamefont {Banerjee}\ \emph {et~al.}(2020)\citenamefont {Banerjee}, \citenamefont {Budker}, \citenamefont {Eby}, \citenamefont {Flambaum}, \citenamefont {Kim}, \citenamefont {Matsedonskyi},\ and\ \citenamefont {Perez}}]{Banerjee:2019epw}%
  \BibitemOpen
  \bibfield  {author} {\bibinfo {author} {\bibfnamefont {A.}~\bibnamefont {Banerjee}}, \bibinfo {author} {\bibfnamefont {D.}~\bibnamefont {Budker}}, \bibinfo {author} {\bibfnamefont {J.}~\bibnamefont {Eby}}, \bibinfo {author} {\bibfnamefont {V.~V.}\ \bibnamefont {Flambaum}}, \bibinfo {author} {\bibfnamefont {H.}~\bibnamefont {Kim}}, \bibinfo {author} {\bibfnamefont {O.}~\bibnamefont {Matsedonskyi}},\ and\ \bibinfo {author} {\bibfnamefont {G.}~\bibnamefont {Perez}},\ }\href {https://doi.org/10.1007/JHEP09(2020)004} {\bibfield  {journal} {\bibinfo  {journal} {{JHEP}}\ }\textbf {\bibinfo {volume} {09}},\ \bibinfo {pages} {004} (\bibinfo {year} {2020})},\ \Eprint {https://arxiv.org/abs/1912.04295} {arXiv:1912.04295} \BibitemShut {NoStop}%
\bibitem [{\citenamefont {Baumann}\ \emph {et~al.}(2019)\citenamefont {Baumann}, \citenamefont {Chia},\ and\ \citenamefont {Porto}}]{Baumann:2018vus}%
  \BibitemOpen
  \bibfield  {author} {\bibinfo {author} {\bibfnamefont {D.}~\bibnamefont {Baumann}}, \bibinfo {author} {\bibfnamefont {H.~S.}\ \bibnamefont {Chia}},\ and\ \bibinfo {author} {\bibfnamefont {R.~A.}\ \bibnamefont {Porto}},\ }\href {https://doi.org/10.1103/PhysRevD.99.044001} {\bibfield  {journal} {\bibinfo  {journal} {Phys. Rev. D}\ }\textbf {\bibinfo {volume} {99}},\ \bibinfo {pages} {044001} (\bibinfo {year} {2019})},\ \Eprint {https://arxiv.org/abs/1804.03208} {arXiv:1804.03208} \BibitemShut {NoStop}%
\bibitem [{\citenamefont {Budker}\ \emph {et~al.}(2023)\citenamefont {Budker}, \citenamefont {Eby}, \citenamefont {Gorghetto}, \citenamefont {Jiang},\ and\ \citenamefont {Perez}}]{Budker:2023sex}%
  \BibitemOpen
  \bibfield  {author} {\bibinfo {author} {\bibfnamefont {D.}~\bibnamefont {Budker}}, \bibinfo {author} {\bibfnamefont {J.}~\bibnamefont {Eby}}, \bibinfo {author} {\bibfnamefont {M.}~\bibnamefont {Gorghetto}}, \bibinfo {author} {\bibfnamefont {M.}~\bibnamefont {Jiang}},\ and\ \bibinfo {author} {\bibfnamefont {G.}~\bibnamefont {Perez}},\ }\href {https://doi.org/10.1088/1475-7516/2023/12/021} {\bibfield  {journal} {\bibinfo  {journal} {{JCAP}}\ }\textbf {\bibinfo {volume} {12}},\ \bibinfo {pages} {021} (\bibinfo {year} {2023})},\ \Eprint {https://arxiv.org/abs/2306.12477} {arXiv:2306.12477 [hep-ph]} \BibitemShut {NoStop}%
\bibitem [{\citenamefont {Levkov}\ \emph {et~al.}(2017)\citenamefont {Levkov}, \citenamefont {Panin},\ and\ \citenamefont {Tkachev}}]{Levkov:2016rkk}%
  \BibitemOpen
  \bibfield  {author} {\bibinfo {author} {\bibfnamefont {D.~G.}\ \bibnamefont {Levkov}}, \bibinfo {author} {\bibfnamefont {A.~G.}\ \bibnamefont {Panin}},\ and\ \bibinfo {author} {\bibfnamefont {I.~I.}\ \bibnamefont {Tkachev}},\ }\href {https://doi.org/10.1103/PhysRevLett.118.011301} {\bibfield  {journal} {\bibinfo  {journal} {Phys. Rev. Lett.}\ }\textbf {\bibinfo {volume} {118}},\ \bibinfo {pages} {011301} (\bibinfo {year} {2017})},\ \Eprint {https://arxiv.org/abs/1609.03611} {arXiv:1609.03611} \BibitemShut {NoStop}%
\bibitem [{\citenamefont {Eby}\ \emph {et~al.}(2016)\citenamefont {Eby}, \citenamefont {Leembruggen}, \citenamefont {Suranyi},\ and\ \citenamefont {Wijewardhana}}]{Eby:2016cnq}%
  \BibitemOpen
  \bibfield  {author} {\bibinfo {author} {\bibfnamefont {J.}~\bibnamefont {Eby}}, \bibinfo {author} {\bibfnamefont {M.}~\bibnamefont {Leembruggen}}, \bibinfo {author} {\bibfnamefont {P.}~\bibnamefont {Suranyi}},\ and\ \bibinfo {author} {\bibfnamefont {L.~C.~R.}\ \bibnamefont {Wijewardhana}},\ }\href {https://doi.org/10.1007/JHEP12(2016)066} {\bibfield  {journal} {\bibinfo  {journal} {{JHEP}}\ }\textbf {\bibinfo {volume} {12}},\ \bibinfo {pages} {066} (\bibinfo {year} {2016})},\ \Eprint {https://arxiv.org/abs/1608.06911} {arXiv:1608.06911} \BibitemShut {NoStop}%
\bibitem [{\citenamefont {Jain}\ \emph {et~al.}(2024)\citenamefont {Jain}, \citenamefont {Wanichwecharungruang},\ and\ \citenamefont {Thomas}}]{Jain:2023tsr}%
  \BibitemOpen
  \bibfield  {author} {\bibinfo {author} {\bibfnamefont {M.}~\bibnamefont {Jain}}, \bibinfo {author} {\bibfnamefont {W.}~\bibnamefont {Wanichwecharungruang}},\ and\ \bibinfo {author} {\bibfnamefont {J.}~\bibnamefont {Thomas}},\ }\href {https://doi.org/10.1103/PhysRevD.109.016002} {\bibfield  {journal} {\bibinfo  {journal} {Phys. Rev. D}\ }\textbf {\bibinfo {volume} {109}},\ \bibinfo {pages} {016002} (\bibinfo {year} {2024})},\ \Eprint {https://arxiv.org/abs/2310.00058} {arXiv:2310.00058} \BibitemShut {NoStop}%
\bibitem [{\citenamefont {Liebling}\ and\ \citenamefont {Palenzuela}(2023)}]{Liebling:2012fv}%
  \BibitemOpen
  \bibfield  {author} {\bibinfo {author} {\bibfnamefont {S.~L.}\ \bibnamefont {Liebling}}\ and\ \bibinfo {author} {\bibfnamefont {C.}~\bibnamefont {Palenzuela}},\ }\href {https://doi.org/10.1007/s41114-023-00043-4} {\bibfield  {journal} {\bibinfo  {journal} {Living Rev. Rel.}\ }\textbf {\bibinfo {volume} {26}},\ \bibinfo {pages} {1} (\bibinfo {year} {2023})},\ \Eprint {https://arxiv.org/abs/1202.5809} {arXiv:1202.5809 [gr-qc]} \BibitemShut {NoStop}%
\bibitem [{\citenamefont {Fox}\ \emph {et~al.}(2023)\citenamefont {Fox}, \citenamefont {Weiner},\ and\ \citenamefont {Xiao}}]{Fox:2023xgx}%
  \BibitemOpen
  \bibfield  {author} {\bibinfo {author} {\bibfnamefont {P.~J.}\ \bibnamefont {Fox}}, \bibinfo {author} {\bibfnamefont {N.}~\bibnamefont {Weiner}},\ and\ \bibinfo {author} {\bibfnamefont {H.}~\bibnamefont {Xiao}},\ }\href {https://doi.org/10.1103/PhysRevD.108.095043} {\bibfield  {journal} {\bibinfo  {journal} {Phys. Rev. D}\ }\textbf {\bibinfo {volume} {108}},\ \bibinfo {pages} {095043} (\bibinfo {year} {2023})},\ \Eprint {https://arxiv.org/abs/2302.00685} {arXiv:2302.00685} \BibitemShut {NoStop}%
\bibitem [{\citenamefont {Maseizik}\ \emph {et~al.}(2025)\citenamefont {Maseizik}, \citenamefont {Eby}, \citenamefont {Seong},\ and\ \citenamefont {Sigl}}]{Maseizik:2024qya}%
  \BibitemOpen
  \bibfield  {author} {\bibinfo {author} {\bibfnamefont {D.}~\bibnamefont {Maseizik}}, \bibinfo {author} {\bibfnamefont {J.}~\bibnamefont {Eby}}, \bibinfo {author} {\bibfnamefont {H.}~\bibnamefont {Seong}},\ and\ \bibinfo {author} {\bibfnamefont {G.}~\bibnamefont {Sigl}},\ }\href {https://doi.org/10.1103/PhysRevD.111.063017} {\bibfield  {journal} {\bibinfo  {journal} {Phys. Rev. D}\ }\textbf {\bibinfo {volume} {111}},\ \bibinfo {pages} {063017} (\bibinfo {year} {2025})},\ \Eprint {https://arxiv.org/abs/2410.13082} {arXiv:2410.13082} \BibitemShut {NoStop}%
\bibitem [{\citenamefont {Chan}\ \emph {et~al.}(2024)\citenamefont {Chan}, \citenamefont {Sibiryakov},\ and\ \citenamefont {Xue}}]{Chan:2022bkz}%
  \BibitemOpen
  \bibfield  {author} {\bibinfo {author} {\bibfnamefont {J.~H.-H.}\ \bibnamefont {Chan}}, \bibinfo {author} {\bibfnamefont {S.}~\bibnamefont {Sibiryakov}},\ and\ \bibinfo {author} {\bibfnamefont {W.}~\bibnamefont {Xue}},\ }\href {https://doi.org/10.1007/JHEP01(2024)071} {\bibfield  {journal} {\bibinfo  {journal} {{JHEP}}\ }\textbf {\bibinfo {volume} {01}},\ \bibinfo {pages} {071} (\bibinfo {year} {2024})},\ \Eprint {https://arxiv.org/abs/2207.04057} {arXiv:2207.04057} \BibitemShut {NoStop}%
\bibitem [{\citenamefont {Wang}\ and\ \citenamefont {Gao}(2026)}]{Wang:2025vhy}%
  \BibitemOpen
  \bibfield  {author} {\bibinfo {author} {\bibfnamefont {Z.}~\bibnamefont {Wang}}\ and\ \bibinfo {author} {\bibfnamefont {Y.}~\bibnamefont {Gao}},\ }\href {https://doi.org/10.1016/j.dark.2026.102290} {\bibfield  {journal} {\bibinfo  {journal} {Phys. Dark Univ.}\ }\textbf {\bibinfo {volume} {52}},\ \bibinfo {pages} {102290} (\bibinfo {year} {2026})},\ \Eprint {https://arxiv.org/abs/2508.14535} {arXiv:2508.14535} \BibitemShut {NoStop}%
\bibitem [{\citenamefont {Chavanis}(2011)}]{Chavanis:2011zi}%
  \BibitemOpen
  \bibfield  {author} {\bibinfo {author} {\bibfnamefont {P.-H.}\ \bibnamefont {Chavanis}},\ }\href {https://doi.org/10.1103/PhysRevD.84.043531} {\bibfield  {journal} {\bibinfo  {journal} {Phys. Rev. D}\ }\textbf {\bibinfo {volume} {84}},\ \bibinfo {pages} {043531} (\bibinfo {year} {2011})},\ \Eprint {https://arxiv.org/abs/1103.2050} {arXiv:1103.2050} \BibitemShut {NoStop}%
\bibitem [{\citenamefont {Schiappacasse}\ and\ \citenamefont {Hertzberg}(2018)}]{Schiappacasse:2017ham}%
  \BibitemOpen
  \bibfield  {author} {\bibinfo {author} {\bibfnamefont {E.~D.}\ \bibnamefont {Schiappacasse}}\ and\ \bibinfo {author} {\bibfnamefont {M.~P.}\ \bibnamefont {Hertzberg}},\ }\href {https://doi.org/10.1088/1475-7516/2018/01/037} {\bibfield  {journal} {\bibinfo  {journal} {{JCAP}}\ }\textbf {\bibinfo {volume} {01}},\ \bibinfo {pages} {037} (\bibinfo {year} {2018})},\ \bibinfo {note} {[Erratum: JCAP 03, E01 (2018)]},\ \Eprint {https://arxiv.org/abs/1710.04729} {arXiv:1710.04729 [hep-ph]} \BibitemShut {NoStop}%
\bibitem [{\citenamefont {van Dokkum}\ \emph {et~al.}(2018)\citenamefont {van Dokkum} \emph {et~al.}}]{vanDokkum:2018vup}%
  \BibitemOpen
  \bibfield  {author} {\bibinfo {author} {\bibfnamefont {P.}~\bibnamefont {van Dokkum}} \emph {et~al.},\ }\href {https://doi.org/10.1038/nature25767} {\bibfield  {journal} {\bibinfo  {journal} {Nature}\ }\textbf {\bibinfo {volume} {555}},\ \bibinfo {pages} {629} (\bibinfo {year} {2018})},\ \Eprint {https://arxiv.org/abs/1803.10237} {arXiv:1803.10237} \BibitemShut {NoStop}%
\bibitem [{\citenamefont {Comer\'on}\ \emph {et~al.}(2023)\citenamefont {Comer\'on}, \citenamefont {Trujillo}, \citenamefont {Cappellari}, \citenamefont {Buitrago}, \citenamefont {Gardu\~no}, \citenamefont {Zaragoza-Cardiel}, \citenamefont {Zinchenko}, \citenamefont {Lara-L\'opez}, \citenamefont {Ferr\'e-Mateu},\ and\ \citenamefont {Dib}}]{Comeron:2023}%
  \BibitemOpen
  \bibfield  {author} {\bibinfo {author} {\bibfnamefont {S.}~\bibnamefont {Comer\'on}}, \bibinfo {author} {\bibfnamefont {I.}~\bibnamefont {Trujillo}}, \bibinfo {author} {\bibfnamefont {M.}~\bibnamefont {Cappellari}}, \bibinfo {author} {\bibfnamefont {F.}~\bibnamefont {Buitrago}}, \bibinfo {author} {\bibfnamefont {L.~E.}\ \bibnamefont {Gardu\~no}}, \bibinfo {author} {\bibfnamefont {J.}~\bibnamefont {Zaragoza-Cardiel}}, \bibinfo {author} {\bibfnamefont {I.~A.}\ \bibnamefont {Zinchenko}}, \bibinfo {author} {\bibfnamefont {M.~A.}\ \bibnamefont {Lara-L\'opez}}, \bibinfo {author} {\bibfnamefont {A.}~\bibnamefont {Ferr\'e-Mateu}},\ and\ \bibinfo {author} {\bibfnamefont {S.}~\bibnamefont {Dib}},\ }\href {https://doi.org/10.1051/0004-6361/202346291} {\bibfield  {journal} {\bibinfo  {journal} {Astron. Astrophys.}\ }\textbf {\bibinfo {volume} {675}},\ \bibinfo {pages} {A143} (\bibinfo {year} {2023})},\ \Eprint {https://arxiv.org/abs/2303.11360} {arXiv:2303.11360} \BibitemShut {NoStop}%
\bibitem [{\citenamefont {Rogers}\ and\ \citenamefont {Peiris}(2021)}]{Rogers:2020ltq}%
  \BibitemOpen
  \bibfield  {author} {\bibinfo {author} {\bibfnamefont {K.~K.}\ \bibnamefont {Rogers}}\ and\ \bibinfo {author} {\bibfnamefont {H.~V.}\ \bibnamefont {Peiris}},\ }\href {https://doi.org/10.1103/PhysRevLett.126.071302} {\bibfield  {journal} {\bibinfo  {journal} {Phys. Rev. Lett.}\ }\textbf {\bibinfo {volume} {126}},\ \bibinfo {pages} {071302} (\bibinfo {year} {2021})},\ \Eprint {https://arxiv.org/abs/2007.12705} {arXiv:2007.12705} \BibitemShut {NoStop}%
\bibitem [{\citenamefont {Co}\ \emph {et~al.}(2020)\citenamefont {Co}, \citenamefont {Hall},\ and\ \citenamefont {Harigaya}}]{Co:2019jts}%
  \BibitemOpen
  \bibfield  {author} {\bibinfo {author} {\bibfnamefont {R.~T.}\ \bibnamefont {Co}}, \bibinfo {author} {\bibfnamefont {L.~J.}\ \bibnamefont {Hall}},\ and\ \bibinfo {author} {\bibfnamefont {K.}~\bibnamefont {Harigaya}},\ }\href {https://doi.org/10.1103/PhysRevLett.124.251802} {\bibfield  {journal} {\bibinfo  {journal} {Phys. Rev. Lett.}\ }\textbf {\bibinfo {volume} {124}},\ \bibinfo {pages} {251802} (\bibinfo {year} {2020})},\ \Eprint {https://arxiv.org/abs/1910.14152} {arXiv:1910.14152} \BibitemShut {NoStop}%
\bibitem [{\citenamefont {Navarro}\ \emph {et~al.}(1997)\citenamefont {Navarro}, \citenamefont {Frenk},\ and\ \citenamefont {White}}]{Navarro:1996gj}%
  \BibitemOpen
  \bibfield  {author} {\bibinfo {author} {\bibfnamefont {J.~F.}\ \bibnamefont {Navarro}}, \bibinfo {author} {\bibfnamefont {C.~S.}\ \bibnamefont {Frenk}},\ and\ \bibinfo {author} {\bibfnamefont {S.~D.~M.}\ \bibnamefont {White}},\ }\href {https://doi.org/10.1086/304888} {\bibfield  {journal} {\bibinfo  {journal} {Astrophys. J.}\ }\textbf {\bibinfo {volume} {490}},\ \bibinfo {pages} {493} (\bibinfo {year} {1997})},\ \Eprint {https://arxiv.org/abs/astro-ph/9611107} {arXiv:astro-ph/9611107} \BibitemShut {NoStop}%
\bibitem [{\citenamefont {McMillan}(2017)}]{McMillan:2016jtx}%
  \BibitemOpen
  \bibfield  {author} {\bibinfo {author} {\bibfnamefont {P.~J.}\ \bibnamefont {McMillan}},\ }\href {https://doi.org/10.1093/mnras/stw2759} {\bibfield  {journal} {\bibinfo  {journal} {Mon. Not. Roy. Astron. Soc.}\ }\textbf {\bibinfo {volume} {465}},\ \bibinfo {pages} {76} (\bibinfo {year} {2017})},\ \Eprint {https://arxiv.org/abs/1608.00971} {arXiv:1608.00971} \BibitemShut {NoStop}%
\bibitem [{\citenamefont {Snaith}\ \emph {et~al.}(2015)\citenamefont {Snaith}, \citenamefont {Haywood}, \citenamefont {Di~Matteo}, \citenamefont {Lehnert}, \citenamefont {Combes}, \citenamefont {Katz},\ and\ \citenamefont {G\'omez}}]{Snaith:2014qxa}%
  \BibitemOpen
  \bibfield  {author} {\bibinfo {author} {\bibfnamefont {O.}~\bibnamefont {Snaith}}, \bibinfo {author} {\bibfnamefont {M.}~\bibnamefont {Haywood}}, \bibinfo {author} {\bibfnamefont {P.}~\bibnamefont {Di~Matteo}}, \bibinfo {author} {\bibfnamefont {M.~D.}\ \bibnamefont {Lehnert}}, \bibinfo {author} {\bibfnamefont {F.}~\bibnamefont {Combes}}, \bibinfo {author} {\bibfnamefont {D.}~\bibnamefont {Katz}},\ and\ \bibinfo {author} {\bibfnamefont {A.}~\bibnamefont {G\'omez}},\ }\href {https://doi.org/10.1051/0004-6361/201424281} {\bibfield  {journal} {\bibinfo  {journal} {Astron. Astrophys.}\ }\textbf {\bibinfo {volume} {578}},\ \bibinfo {pages} {A87} (\bibinfo {year} {2015})},\ \Eprint {https://arxiv.org/abs/1410.3829} {arXiv:1410.3829} \BibitemShut {NoStop}%
\bibitem [{\citenamefont {Hernquist}(1990)}]{Hernquist:1990be}%
  \BibitemOpen
  \bibfield  {author} {\bibinfo {author} {\bibfnamefont {L.}~\bibnamefont {Hernquist}},\ }\href {https://doi.org/10.1086/168845} {\bibfield  {journal} {\bibinfo  {journal} {Astrophys. J.}\ }\textbf {\bibinfo {volume} {356}},\ \bibinfo {pages} {359} (\bibinfo {year} {1990})}\BibitemShut {NoStop}%
\bibitem [{\citenamefont {Kroupa}(2001)}]{Kroupa:2000iv}%
  \BibitemOpen
  \bibfield  {author} {\bibinfo {author} {\bibfnamefont {P.}~\bibnamefont {Kroupa}},\ }\href {https://doi.org/10.1046/j.1365-8711.2001.04022.x} {\bibfield  {journal} {\bibinfo  {journal} {Mon. Not. Roy. Astron. Soc.}\ }\textbf {\bibinfo {volume} {322}},\ \bibinfo {pages} {231} (\bibinfo {year} {2001})},\ \Eprint {https://arxiv.org/abs/astro-ph/0009005} {arXiv:astro-ph/0009005} \BibitemShut {NoStop}%
\bibitem [{\citenamefont {Leitner}\ and\ \citenamefont {Kravtsov}(2011)}]{Leitner:2010nh}%
  \BibitemOpen
  \bibfield  {author} {\bibinfo {author} {\bibfnamefont {S.~N.}\ \bibnamefont {Leitner}}\ and\ \bibinfo {author} {\bibfnamefont {A.~V.}\ \bibnamefont {Kravtsov}},\ }\href {https://doi.org/10.1088/0004-637X/734/1/48} {\bibfield  {journal} {\bibinfo  {journal} {Astrophys. J.}\ }\textbf {\bibinfo {volume} {734}},\ \bibinfo {pages} {48} (\bibinfo {year} {2011})},\ \Eprint {https://arxiv.org/abs/1011.1252} {arXiv:1011.1252} \BibitemShut {NoStop}%
\bibitem [{\citenamefont {Hurley}\ \emph {et~al.}(2000)\citenamefont {Hurley}, \citenamefont {Pols},\ and\ \citenamefont {Tout}}]{Hurley:2000pk}%
  \BibitemOpen
  \bibfield  {author} {\bibinfo {author} {\bibfnamefont {J.~R.}\ \bibnamefont {Hurley}}, \bibinfo {author} {\bibfnamefont {O.~R.}\ \bibnamefont {Pols}},\ and\ \bibinfo {author} {\bibfnamefont {C.~A.}\ \bibnamefont {Tout}},\ }\href {https://doi.org/10.1046/j.1365-8711.2000.03426.x} {\bibfield  {journal} {\bibinfo  {journal} {Mon. Not. Roy. Astron. Soc.}\ }\textbf {\bibinfo {volume} {315}},\ \bibinfo {pages} {543} (\bibinfo {year} {2000})},\ \Eprint {https://arxiv.org/abs/astro-ph/0001295} {arXiv:astro-ph/0001295} \BibitemShut {NoStop}%
\bibitem [{\citenamefont {Kalirai}\ \emph {et~al.}(2008)\citenamefont {Kalirai}, \citenamefont {Hansen}, \citenamefont {Kelson}, \citenamefont {Reitzel}, \citenamefont {Rich},\ and\ \citenamefont {Richer}}]{Kalirai:2008}%
  \BibitemOpen
  \bibfield  {author} {\bibinfo {author} {\bibfnamefont {J.~S.}\ \bibnamefont {Kalirai}}, \bibinfo {author} {\bibfnamefont {B.~M.~S.}\ \bibnamefont {Hansen}}, \bibinfo {author} {\bibfnamefont {D.~D.}\ \bibnamefont {Kelson}}, \bibinfo {author} {\bibfnamefont {D.~B.}\ \bibnamefont {Reitzel}}, \bibinfo {author} {\bibfnamefont {R.~M.}\ \bibnamefont {Rich}},\ and\ \bibinfo {author} {\bibfnamefont {H.~B.}\ \bibnamefont {Richer}},\ }\href {https://doi.org/10.1086/527028} {\bibfield  {journal} {\bibinfo  {journal} {Astrophys. J.}\ }\textbf {\bibinfo {volume} {676}},\ \bibinfo {pages} {594} (\bibinfo {year} {2008})},\ \Eprint {https://arxiv.org/abs/0706.3894} {arXiv:0706.3894} \BibitemShut {NoStop}%
\bibitem [{\citenamefont {Fryer}\ \emph {et~al.}(2012)\citenamefont {Fryer}, \citenamefont {Belczynski}, \citenamefont {Wiktorowicz}, \citenamefont {Dominik}, \citenamefont {Kalogera},\ and\ \citenamefont {Holz}}]{Fryer:2011cx}%
  \BibitemOpen
  \bibfield  {author} {\bibinfo {author} {\bibfnamefont {C.~L.}\ \bibnamefont {Fryer}}, \bibinfo {author} {\bibfnamefont {K.}~\bibnamefont {Belczynski}}, \bibinfo {author} {\bibfnamefont {G.}~\bibnamefont {Wiktorowicz}}, \bibinfo {author} {\bibfnamefont {M.}~\bibnamefont {Dominik}}, \bibinfo {author} {\bibfnamefont {V.}~\bibnamefont {Kalogera}},\ and\ \bibinfo {author} {\bibfnamefont {D.~E.}\ \bibnamefont {Holz}},\ }\href {https://doi.org/10.1088/0004-637X/749/1/91} {\bibfield  {journal} {\bibinfo  {journal} {Astrophys. J.}\ }\textbf {\bibinfo {volume} {749}},\ \bibinfo {pages} {91} (\bibinfo {year} {2012})},\ \Eprint {https://arxiv.org/abs/1110.1726} {arXiv:1110.1726} \BibitemShut {NoStop}%
\bibitem [{\citenamefont {Sartore}\ \emph {et~al.}(2010)\citenamefont {Sartore}, \citenamefont {Ripamonti}, \citenamefont {Treves},\ and\ \citenamefont {Turolla}}]{Sartore:2009wn}%
  \BibitemOpen
  \bibfield  {author} {\bibinfo {author} {\bibfnamefont {N.}~\bibnamefont {Sartore}}, \bibinfo {author} {\bibfnamefont {E.}~\bibnamefont {Ripamonti}}, \bibinfo {author} {\bibfnamefont {A.}~\bibnamefont {Treves}},\ and\ \bibinfo {author} {\bibfnamefont {R.}~\bibnamefont {Turolla}},\ }\href {https://doi.org/10.1051/0004-6361/200912222} {\bibfield  {journal} {\bibinfo  {journal} {Astron. Astrophys.}\ }\textbf {\bibinfo {volume} {510}},\ \bibinfo {pages} {A23} (\bibinfo {year} {2010})},\ \Eprint {https://arxiv.org/abs/0908.3182} {arXiv:0908.3182} \BibitemShut {NoStop}%
\bibitem [{\citenamefont {Abuter}\ \emph {et~al.}(2018)\citenamefont {Abuter} \emph {et~al.}}]{GRAVITY:2018ofz}%
  \BibitemOpen
  \bibfield  {author} {\bibinfo {author} {\bibfnamefont {R.}~\bibnamefont {Abuter}} \emph {et~al.} (\bibinfo {collaboration} {GRAVITY}),\ }\href {https://doi.org/10.1051/0004-6361/201833718} {\bibfield  {journal} {\bibinfo  {journal} {Astron. Astrophys.}\ }\textbf {\bibinfo {volume} {615}},\ \bibinfo {pages} {L15} (\bibinfo {year} {2018})},\ \Eprint {https://arxiv.org/abs/1807.09409} {arXiv:1807.09409} \BibitemShut {NoStop}%
\bibitem [{\citenamefont {Dalal}\ and\ \citenamefont {Kravtsov}(2022)}]{Dalal:2022rmp}%
  \BibitemOpen
  \bibfield  {author} {\bibinfo {author} {\bibfnamefont {N.}~\bibnamefont {Dalal}}\ and\ \bibinfo {author} {\bibfnamefont {A.}~\bibnamefont {Kravtsov}},\ }\href {https://doi.org/10.1103/PhysRevD.106.063517} {\bibfield  {journal} {\bibinfo  {journal} {Phys. Rev. D}\ }\textbf {\bibinfo {volume} {106}},\ \bibinfo {pages} {063517} (\bibinfo {year} {2022})},\ \Eprint {https://arxiv.org/abs/2203.05750} {arXiv:2203.05750} \BibitemShut {NoStop}%
\bibitem [{\citenamefont {Bar}\ \emph {et~al.}(2019)\citenamefont {Bar}, \citenamefont {Blum}, \citenamefont {Lacroix},\ and\ \citenamefont {Panci}}]{Bar:2019pnz}%
  \BibitemOpen
  \bibfield  {author} {\bibinfo {author} {\bibfnamefont {N.}~\bibnamefont {Bar}}, \bibinfo {author} {\bibfnamefont {K.}~\bibnamefont {Blum}}, \bibinfo {author} {\bibfnamefont {T.}~\bibnamefont {Lacroix}},\ and\ \bibinfo {author} {\bibfnamefont {P.}~\bibnamefont {Panci}},\ }\href {https://doi.org/10.1088/1475-7516/2019/07/045} {\bibfield  {journal} {\bibinfo  {journal} {{JCAP}}\ }\textbf {\bibinfo {volume} {07}},\ \bibinfo {pages} {045} (\bibinfo {year} {2019})},\ \Eprint {https://arxiv.org/abs/1905.11745} {arXiv:1905.11745} \BibitemShut {NoStop}%
\bibitem [{\citenamefont {Burkert}(1995)}]{Burkert:1995yz}%
  \BibitemOpen
  \bibfield  {author} {\bibinfo {author} {\bibfnamefont {A.}~\bibnamefont {Burkert}},\ }\href {https://doi.org/10.1086/309560} {\bibfield  {journal} {\bibinfo  {journal} {Astrophys. J. Lett.}\ }\textbf {\bibinfo {volume} {447}},\ \bibinfo {pages} {L25} (\bibinfo {year} {1995})},\ \Eprint {https://arxiv.org/abs/astro-ph/9504041} {arXiv:astro-ph/9504041} \BibitemShut {NoStop}%
\bibitem [{\citenamefont {Salucci}\ \emph {et~al.}(2012)\citenamefont {Salucci}, \citenamefont {Wilkinson}, \citenamefont {Walker}, \citenamefont {Gilmore}, \citenamefont {Grebel}, \citenamefont {Koch}, \citenamefont {Martins},\ and\ \citenamefont {Wyse}}]{Salucci:2011ee}%
  \BibitemOpen
  \bibfield  {author} {\bibinfo {author} {\bibfnamefont {P.}~\bibnamefont {Salucci}}, \bibinfo {author} {\bibfnamefont {M.~I.}\ \bibnamefont {Wilkinson}}, \bibinfo {author} {\bibfnamefont {M.~G.}\ \bibnamefont {Walker}}, \bibinfo {author} {\bibfnamefont {G.~F.}\ \bibnamefont {Gilmore}}, \bibinfo {author} {\bibfnamefont {E.~K.}\ \bibnamefont {Grebel}}, \bibinfo {author} {\bibfnamefont {A.}~\bibnamefont {Koch}}, \bibinfo {author} {\bibfnamefont {C.~F.}\ \bibnamefont {Martins}},\ and\ \bibinfo {author} {\bibfnamefont {R.~F.~G.}\ \bibnamefont {Wyse}},\ }\href {https://doi.org/10.1111/j.1365-2966.2011.20144.x} {\bibfield  {journal} {\bibinfo  {journal} {Mon. Not. Roy. Astron. Soc.}\ }\textbf {\bibinfo {volume} {420}},\ \bibinfo {pages} {2034} (\bibinfo {year} {2012})},\ \Eprint {https://arxiv.org/abs/1111.1165} {arXiv:1111.1165 [astro-ph.CO]} \BibitemShut {NoStop}%
\bibitem [{\citenamefont {de~Boer}\ \emph {et~al.}(2012)\citenamefont {de~Boer}, \citenamefont {Tolstoy}, \citenamefont {Hill}, \citenamefont {Saha}, \citenamefont {Olszewski}, \citenamefont {Mateo}, \citenamefont {Starkenburg}, \citenamefont {Battaglia},\ and\ \citenamefont {Walker}}]{deBoer:2012py}%
  \BibitemOpen
  \bibfield  {author} {\bibinfo {author} {\bibfnamefont {T.~J.~L.}\ \bibnamefont {de~Boer}}, \bibinfo {author} {\bibfnamefont {E.}~\bibnamefont {Tolstoy}}, \bibinfo {author} {\bibfnamefont {V.}~\bibnamefont {Hill}}, \bibinfo {author} {\bibfnamefont {A.}~\bibnamefont {Saha}}, \bibinfo {author} {\bibfnamefont {E.~W.}\ \bibnamefont {Olszewski}}, \bibinfo {author} {\bibfnamefont {M.}~\bibnamefont {Mateo}}, \bibinfo {author} {\bibfnamefont {E.}~\bibnamefont {Starkenburg}}, \bibinfo {author} {\bibfnamefont {G.}~\bibnamefont {Battaglia}},\ and\ \bibinfo {author} {\bibfnamefont {M.~G.}\ \bibnamefont {Walker}},\ }\href {https://doi.org/10.1051/0004-6361/201219547} {\bibfield  {journal} {\bibinfo  {journal} {Astron. Astrophys.}\ }\textbf {\bibinfo {volume} {544}},\ \bibinfo {pages} {A73} (\bibinfo {year} {2012})},\ \Eprint {https://arxiv.org/abs/1206.6968} {arXiv:1206.6968 [astro-ph.GA]} \BibitemShut {NoStop}%
\bibitem [{\citenamefont {McConnachie}(2012)}]{McConnachie:2012vd}%
  \BibitemOpen
  \bibfield  {author} {\bibinfo {author} {\bibfnamefont {A.~W.}\ \bibnamefont {McConnachie}},\ }\href {https://doi.org/10.1088/0004-6256/144/1/4} {\bibfield  {journal} {\bibinfo  {journal} {Astron. J.}\ }\textbf {\bibinfo {volume} {144}},\ \bibinfo {pages} {4} (\bibinfo {year} {2012})},\ \Eprint {https://arxiv.org/abs/1204.1562} {arXiv:1204.1562 [astro-ph.CO]} \BibitemShut {NoStop}%
\bibitem [{\citenamefont {Fensch}\ \emph {et~al.}(2019)\citenamefont {Fensch}, \citenamefont {van~der Burg}, \citenamefont {Je{\v{r}}{\'a}bkov{\'a}}, \citenamefont {Emsellem}, \citenamefont {Zanella}, \citenamefont {Agnello}, \citenamefont {Hilker}, \citenamefont {M{\"u}ller}, \citenamefont {Rejkuba}, \citenamefont {Duc}, \citenamefont {Durrell}, \citenamefont {Habas}, \citenamefont {Lim}, \citenamefont {Marleau}, \citenamefont {Peng},\ and\ \citenamefont {S{\'a}nchez~Janssen}}]{Fensch:2019}%
  \BibitemOpen
  \bibfield  {author} {\bibinfo {author} {\bibfnamefont {J.}~\bibnamefont {Fensch}}, \bibinfo {author} {\bibfnamefont {R.~F.~J.}\ \bibnamefont {van~der Burg}}, \bibinfo {author} {\bibfnamefont {T.}~\bibnamefont {Je{\v{r}}{\'a}bkov{\'a}}}, \bibinfo {author} {\bibfnamefont {E.}~\bibnamefont {Emsellem}}, \bibinfo {author} {\bibfnamefont {A.}~\bibnamefont {Zanella}}, \bibinfo {author} {\bibfnamefont {A.}~\bibnamefont {Agnello}}, \bibinfo {author} {\bibfnamefont {M.}~\bibnamefont {Hilker}}, \bibinfo {author} {\bibfnamefont {O.}~\bibnamefont {M{\"u}ller}}, \bibinfo {author} {\bibfnamefont {M.}~\bibnamefont {Rejkuba}}, \bibinfo {author} {\bibfnamefont {P.-A.}\ \bibnamefont {Duc}}, \bibinfo {author} {\bibfnamefont {P.}~\bibnamefont {Durrell}}, \bibinfo {author} {\bibfnamefont {R.}~\bibnamefont {Habas}}, \bibinfo {author} {\bibfnamefont {S.}~\bibnamefont {Lim}}, \bibinfo {author} {\bibfnamefont {F.~R.}\ \bibnamefont {Marleau}}, \bibinfo {author} {\bibfnamefont {E.~W.}\ \bibnamefont {Peng}},\ and\ \bibinfo {author}
  {\bibfnamefont {R.}~\bibnamefont {S{\'a}nchez~Janssen}},\ }\href {https://doi.org/10.1051/0004-6361/201834911} {\bibfield  {journal} {\bibinfo  {journal} {Astron. Astrophys.}\ }\textbf {\bibinfo {volume} {625}},\ \bibinfo {pages} {A77} (\bibinfo {year} {2019})},\ \Eprint {https://arxiv.org/abs/1812.07346} {arXiv:1812.07346 [astro-ph.GA]} \BibitemShut {NoStop}%
\bibitem [{\citenamefont {Danieli}\ \emph {et~al.}(2019)\citenamefont {Danieli}, \citenamefont {van Dokkum}, \citenamefont {Conroy}, \citenamefont {Abraham},\ and\ \citenamefont {Romanowsky}}]{Danieli:2019zyi}%
  \BibitemOpen
  \bibfield  {author} {\bibinfo {author} {\bibfnamefont {S.}~\bibnamefont {Danieli}}, \bibinfo {author} {\bibfnamefont {P.}~\bibnamefont {van Dokkum}}, \bibinfo {author} {\bibfnamefont {C.}~\bibnamefont {Conroy}}, \bibinfo {author} {\bibfnamefont {R.}~\bibnamefont {Abraham}},\ and\ \bibinfo {author} {\bibfnamefont {A.~J.}\ \bibnamefont {Romanowsky}},\ }\href {https://doi.org/10.3847/2041-8213/ab0e8c} {\bibfield  {journal} {\bibinfo  {journal} {Astrophys. J. Lett.}\ }\textbf {\bibinfo {volume} {874}},\ \bibinfo {pages} {L12} (\bibinfo {year} {2019})},\ \Eprint {https://arxiv.org/abs/1901.03711} {arXiv:1901.03711 [astro-ph.GA]} \BibitemShut {NoStop}%
\bibitem [{\citenamefont {Trujillo}\ \emph {et~al.}(2014)\citenamefont {Trujillo}, \citenamefont {Ferr{\'e}-Mateu}, \citenamefont {Balcells}, \citenamefont {Vazdekis},\ and\ \citenamefont {S{\'a}nchez-Bl{\'a}zquez}}]{Trujillo:2013ana}%
  \BibitemOpen
  \bibfield  {author} {\bibinfo {author} {\bibfnamefont {I.}~\bibnamefont {Trujillo}}, \bibinfo {author} {\bibfnamefont {A.}~\bibnamefont {Ferr{\'e}-Mateu}}, \bibinfo {author} {\bibfnamefont {M.}~\bibnamefont {Balcells}}, \bibinfo {author} {\bibfnamefont {A.}~\bibnamefont {Vazdekis}},\ and\ \bibinfo {author} {\bibfnamefont {P.}~\bibnamefont {S{\'a}nchez-Bl{\'a}zquez}},\ }\href {https://doi.org/10.1088/2041-8205/780/2/L20} {\bibfield  {journal} {\bibinfo  {journal} {Astrophys. J. Lett.}\ }\textbf {\bibinfo {volume} {780}},\ \bibinfo {pages} {L20} (\bibinfo {year} {2014})},\ \Eprint {https://arxiv.org/abs/1310.6367} {arXiv:1310.6367 [astro-ph.CO]} \BibitemShut {NoStop}%
\bibitem [{\citenamefont {Y{\i}ld{\i}r{\i}m}\ \emph {et~al.}(2015)\citenamefont {Y{\i}ld{\i}r{\i}m}, \citenamefont {van~den Bosch}, \citenamefont {van~de Ven}, \citenamefont {Husemann}, \citenamefont {Lyubenova}, \citenamefont {Walsh}, \citenamefont {Gebhardt},\ and\ \citenamefont {G{\"u}ltekin}}]{Yildirim:2015zxa}%
  \BibitemOpen
  \bibfield  {author} {\bibinfo {author} {\bibfnamefont {A.}~\bibnamefont {Y{\i}ld{\i}r{\i}m}}, \bibinfo {author} {\bibfnamefont {R.~C.~E.}\ \bibnamefont {van~den Bosch}}, \bibinfo {author} {\bibfnamefont {G.}~\bibnamefont {van~de Ven}}, \bibinfo {author} {\bibfnamefont {B.}~\bibnamefont {Husemann}}, \bibinfo {author} {\bibfnamefont {M.}~\bibnamefont {Lyubenova}}, \bibinfo {author} {\bibfnamefont {J.~L.}\ \bibnamefont {Walsh}}, \bibinfo {author} {\bibfnamefont {K.}~\bibnamefont {Gebhardt}},\ and\ \bibinfo {author} {\bibfnamefont {K.}~\bibnamefont {G{\"u}ltekin}},\ }\href {https://doi.org/10.1093/mnras/stv1381} {\bibfield  {journal} {\bibinfo  {journal} {Mon. Not. Roy. Astron. Soc.}\ }\textbf {\bibinfo {volume} {452}},\ \bibinfo {pages} {1792} (\bibinfo {year} {2015})},\ \Eprint {https://arxiv.org/abs/1506.06762} {arXiv:1506.06762 [astro-ph.GA]} \BibitemShut {NoStop}%
\bibitem [{\citenamefont {Behroozi}\ \emph {et~al.}(2013)\citenamefont {Behroozi}, \citenamefont {Wechsler},\ and\ \citenamefont {Conroy}}]{Behroozi:2012iw}%
  \BibitemOpen
  \bibfield  {author} {\bibinfo {author} {\bibfnamefont {P.~S.}\ \bibnamefont {Behroozi}}, \bibinfo {author} {\bibfnamefont {R.~H.}\ \bibnamefont {Wechsler}},\ and\ \bibinfo {author} {\bibfnamefont {C.}~\bibnamefont {Conroy}},\ }\href {https://doi.org/10.1088/0004-637X/770/1/57} {\bibfield  {journal} {\bibinfo  {journal} {Astrophys. J.}\ }\textbf {\bibinfo {volume} {770}},\ \bibinfo {pages} {57} (\bibinfo {year} {2013})},\ \Eprint {https://arxiv.org/abs/1207.6105} {arXiv:1207.6105 [astro-ph.CO]} \BibitemShut {NoStop}%
\bibitem [{\citenamefont {Dutton}\ and\ \citenamefont {Macci{\`o}}(2014)}]{Dutton:2014xda}%
  \BibitemOpen
  \bibfield  {author} {\bibinfo {author} {\bibfnamefont {A.~A.}\ \bibnamefont {Dutton}}\ and\ \bibinfo {author} {\bibfnamefont {A.~V.}\ \bibnamefont {Macci{\`o}}},\ }\href {https://doi.org/10.1093/mnras/stu742} {\bibfield  {journal} {\bibinfo  {journal} {Mon. Not. Roy. Astron. Soc.}\ }\textbf {\bibinfo {volume} {441}},\ \bibinfo {pages} {3359} (\bibinfo {year} {2014})},\ \Eprint {https://arxiv.org/abs/1402.7073} {arXiv:1402.7073 [astro-ph.CO]} \BibitemShut {NoStop}%
\bibitem [{\citenamefont {Kravtsov}\ \emph {et~al.}(2018)\citenamefont {Kravtsov}, \citenamefont {Vikhlinin},\ and\ \citenamefont {Meshscheryakov}}]{Kravtsov:2014sra}%
  \BibitemOpen
  \bibfield  {author} {\bibinfo {author} {\bibfnamefont {A.}~\bibnamefont {Kravtsov}}, \bibinfo {author} {\bibfnamefont {A.}~\bibnamefont {Vikhlinin}},\ and\ \bibinfo {author} {\bibfnamefont {A.}~\bibnamefont {Meshscheryakov}},\ }\href {https://doi.org/10.1134/S1063773717120015} {\bibfield  {journal} {\bibinfo  {journal} {Astron. Lett.}\ }\textbf {\bibinfo {volume} {44}},\ \bibinfo {pages} {8} (\bibinfo {year} {2018})},\ \Eprint {https://arxiv.org/abs/1401.7329} {arXiv:1401.7329 [astro-ph.CO]} \BibitemShut {NoStop}%
\bibitem [{\citenamefont {Shen}\ \emph {et~al.}(2003)\citenamefont {Shen}, \citenamefont {Mo}, \citenamefont {White}, \citenamefont {Blanton}, \citenamefont {Kauffmann}, \citenamefont {Voges}, \citenamefont {Brinkmann},\ and\ \citenamefont {Csabai}}]{Shen:2003sda}%
  \BibitemOpen
  \bibfield  {author} {\bibinfo {author} {\bibfnamefont {S.}~\bibnamefont {Shen}}, \bibinfo {author} {\bibfnamefont {H.~J.}\ \bibnamefont {Mo}}, \bibinfo {author} {\bibfnamefont {S.~D.~M.}\ \bibnamefont {White}}, \bibinfo {author} {\bibfnamefont {M.~R.}\ \bibnamefont {Blanton}}, \bibinfo {author} {\bibfnamefont {G.}~\bibnamefont {Kauffmann}}, \bibinfo {author} {\bibfnamefont {W.}~\bibnamefont {Voges}}, \bibinfo {author} {\bibfnamefont {J.}~\bibnamefont {Brinkmann}},\ and\ \bibinfo {author} {\bibfnamefont {I.}~\bibnamefont {Csabai}},\ }\href {https://doi.org/10.1046/j.1365-8711.2003.06740.x} {\bibfield  {journal} {\bibinfo  {journal} {Mon. Not. Roy. Astron. Soc.}\ }\textbf {\bibinfo {volume} {343}},\ \bibinfo {pages} {978} (\bibinfo {year} {2003})},\ \bibinfo {note} {[Erratum: Mon. Not. Roy. Astron. Soc. 379, 400 (2007)]},\ \Eprint {https://arxiv.org/abs/astro-ph/0301527} {arXiv:astro-ph/0301527} \BibitemShut {NoStop}%
\bibitem [{\citenamefont {Zahid}\ \emph {et~al.}(2016)\citenamefont {Zahid}, \citenamefont {Geller}, \citenamefont {Fabricant},\ and\ \citenamefont {Hwang}}]{Zahid:2016}%
  \BibitemOpen
  \bibfield  {author} {\bibinfo {author} {\bibfnamefont {H.~J.}\ \bibnamefont {Zahid}}, \bibinfo {author} {\bibfnamefont {M.~J.}\ \bibnamefont {Geller}}, \bibinfo {author} {\bibfnamefont {D.~G.}\ \bibnamefont {Fabricant}},\ and\ \bibinfo {author} {\bibfnamefont {H.~S.}\ \bibnamefont {Hwang}},\ }\href {https://doi.org/10.3847/0004-637X/832/2/203} {\bibfield  {journal} {\bibinfo  {journal} {Astrophys. J.}\ }\textbf {\bibinfo {volume} {832}},\ \bibinfo {pages} {203} (\bibinfo {year} {2016})},\ \Eprint {https://arxiv.org/abs/1607.04275} {arXiv:1607.04275 [astro-ph.GA]} \BibitemShut {NoStop}%
\bibitem [{\citenamefont {Thomas}\ \emph {et~al.}(2005)\citenamefont {Thomas}, \citenamefont {Maraston}, \citenamefont {Bender},\ and\ \citenamefont {Mendes~de Oliveira}}]{Thomas:2004ep}%
  \BibitemOpen
  \bibfield  {author} {\bibinfo {author} {\bibfnamefont {D.}~\bibnamefont {Thomas}}, \bibinfo {author} {\bibfnamefont {C.}~\bibnamefont {Maraston}}, \bibinfo {author} {\bibfnamefont {R.}~\bibnamefont {Bender}},\ and\ \bibinfo {author} {\bibfnamefont {C.}~\bibnamefont {Mendes~de Oliveira}},\ }\href {https://doi.org/10.1086/426932} {\bibfield  {journal} {\bibinfo  {journal} {Astrophys. J.}\ }\textbf {\bibinfo {volume} {621}},\ \bibinfo {pages} {673} (\bibinfo {year} {2005})},\ \Eprint {https://arxiv.org/abs/astro-ph/0410209} {arXiv:astro-ph/0410209} \BibitemShut {NoStop}%
\bibitem [{\citenamefont {Moore}\ \emph {et~al.}(1999)\citenamefont {Moore}, \citenamefont {Quinn}, \citenamefont {Governato}, \citenamefont {Stadel},\ and\ \citenamefont {Lake}}]{Moore:1999gc}%
  \BibitemOpen
  \bibfield  {author} {\bibinfo {author} {\bibfnamefont {B.}~\bibnamefont {Moore}}, \bibinfo {author} {\bibfnamefont {T.~R.}\ \bibnamefont {Quinn}}, \bibinfo {author} {\bibfnamefont {F.}~\bibnamefont {Governato}}, \bibinfo {author} {\bibfnamefont {J.}~\bibnamefont {Stadel}},\ and\ \bibinfo {author} {\bibfnamefont {G.}~\bibnamefont {Lake}},\ }\href {https://doi.org/10.1046/j.1365-8711.1999.03039.x} {\bibfield  {journal} {\bibinfo  {journal} {Mon. Not. Roy. Astron. Soc.}\ }\textbf {\bibinfo {volume} {310}},\ \bibinfo {pages} {1147} (\bibinfo {year} {1999})},\ \Eprint {https://arxiv.org/abs/astro-ph/9903164} {arXiv:astro-ph/9903164} \BibitemShut {NoStop}%
\bibitem [{\citenamefont {Sch{\"o}del}\ \emph {et~al.}(2014)\citenamefont {Sch{\"o}del}, \citenamefont {Feldmeier}, \citenamefont {Kunneriath}, \citenamefont {Stolovy}, \citenamefont {Neumayer}, \citenamefont {Amaro-Seoane},\ and\ \citenamefont {Nishiyama}}]{Schodel:2014gna}%
  \BibitemOpen
  \bibfield  {author} {\bibinfo {author} {\bibfnamefont {R.}~\bibnamefont {Sch{\"o}del}}, \bibinfo {author} {\bibfnamefont {A.}~\bibnamefont {Feldmeier}}, \bibinfo {author} {\bibfnamefont {D.}~\bibnamefont {Kunneriath}}, \bibinfo {author} {\bibfnamefont {S.}~\bibnamefont {Stolovy}}, \bibinfo {author} {\bibfnamefont {N.}~\bibnamefont {Neumayer}}, \bibinfo {author} {\bibfnamefont {P.}~\bibnamefont {Amaro-Seoane}},\ and\ \bibinfo {author} {\bibfnamefont {S.}~\bibnamefont {Nishiyama}},\ }\href {https://doi.org/10.1051/0004-6361/201423481} {\bibfield  {journal} {\bibinfo  {journal} {Astron. Astrophys.}\ }\textbf {\bibinfo {volume} {566}},\ \bibinfo {pages} {A47} (\bibinfo {year} {2014})},\ \Eprint {https://arxiv.org/abs/1403.6657} {arXiv:1403.6657 [astro-ph.GA]} \BibitemShut {NoStop}%
\bibitem [{\citenamefont {Sch{\"o}del}\ \emph {et~al.}(2020)\citenamefont {Sch{\"o}del}, \citenamefont {Nogueras-Lara}, \citenamefont {Gallego-Cano}, \citenamefont {Shahzamanian}, \citenamefont {Gallego-Calvente},\ and\ \citenamefont {Gardini}}]{Schodel:2020}%
  \BibitemOpen
  \bibfield  {author} {\bibinfo {author} {\bibfnamefont {R.}~\bibnamefont {Sch{\"o}del}}, \bibinfo {author} {\bibfnamefont {F.}~\bibnamefont {Nogueras-Lara}}, \bibinfo {author} {\bibfnamefont {E.}~\bibnamefont {Gallego-Cano}}, \bibinfo {author} {\bibfnamefont {B.}~\bibnamefont {Shahzamanian}}, \bibinfo {author} {\bibfnamefont {A.~T.}\ \bibnamefont {Gallego-Calvente}},\ and\ \bibinfo {author} {\bibfnamefont {A.}~\bibnamefont {Gardini}},\ }\href {https://doi.org/10.1051/0004-6361/201936688} {\bibfield  {journal} {\bibinfo  {journal} {Astron. Astrophys.}\ }\textbf {\bibinfo {volume} {641}},\ \bibinfo {pages} {A102} (\bibinfo {year} {2020})},\ \Eprint {https://arxiv.org/abs/2007.15950} {arXiv:2007.15950 [astro-ph.GA]} \BibitemShut {NoStop}%
\bibitem [{\citenamefont {Jardel}\ and\ \citenamefont {Gebhardt}(2012)}]{Jardel:2012}%
  \BibitemOpen
  \bibfield  {author} {\bibinfo {author} {\bibfnamefont {J.~R.}\ \bibnamefont {Jardel}}\ and\ \bibinfo {author} {\bibfnamefont {K.}~\bibnamefont {Gebhardt}},\ }\href {https://doi.org/10.1088/0004-637X/746/1/89} {\bibfield  {journal} {\bibinfo  {journal} {Astrophys. J.}\ }\textbf {\bibinfo {volume} {746}},\ \bibinfo {pages} {89} (\bibinfo {year} {2012})},\ \Eprint {https://arxiv.org/abs/1112.0319} {arXiv:1112.0319 [astro-ph.CO]} \BibitemShut {NoStop}%
\bibitem [{\citenamefont {Reines}\ and\ \citenamefont {Volonteri}(2015)}]{Reines:2015}%
  \BibitemOpen
  \bibfield  {author} {\bibinfo {author} {\bibfnamefont {A.~E.}\ \bibnamefont {Reines}}\ and\ \bibinfo {author} {\bibfnamefont {M.}~\bibnamefont {Volonteri}},\ }\href {https://doi.org/10.1088/0004-637X/813/2/82} {\bibfield  {journal} {\bibinfo  {journal} {Astrophys. J.}\ }\textbf {\bibinfo {volume} {813}},\ \bibinfo {pages} {82} (\bibinfo {year} {2015})},\ \Eprint {https://arxiv.org/abs/1508.06274} {arXiv:1508.06274 [astro-ph.GA]} \BibitemShut {NoStop}%
\bibitem [{\citenamefont {Walsh}\ \emph {et~al.}(2016)\citenamefont {Walsh}, \citenamefont {van~den Bosch}, \citenamefont {Gebhardt}, \citenamefont {Y{\i}ld{\i}r{\i}m}, \citenamefont {Richstone}, \citenamefont {G{\"u}ltekin},\ and\ \citenamefont {Husemann}}]{Walsh:2016}%
  \BibitemOpen
  \bibfield  {author} {\bibinfo {author} {\bibfnamefont {J.~L.}\ \bibnamefont {Walsh}}, \bibinfo {author} {\bibfnamefont {R.~C.~E.}\ \bibnamefont {van~den Bosch}}, \bibinfo {author} {\bibfnamefont {K.}~\bibnamefont {Gebhardt}}, \bibinfo {author} {\bibfnamefont {A.}~\bibnamefont {Y{\i}ld{\i}r{\i}m}}, \bibinfo {author} {\bibfnamefont {D.~O.}\ \bibnamefont {Richstone}}, \bibinfo {author} {\bibfnamefont {K.}~\bibnamefont {G{\"u}ltekin}},\ and\ \bibinfo {author} {\bibfnamefont {B.}~\bibnamefont {Husemann}},\ }\href {https://doi.org/10.3847/0004-637X/817/1/2} {\bibfield  {journal} {\bibinfo  {journal} {Astrophys. J.}\ }\textbf {\bibinfo {volume} {817}},\ \bibinfo {pages} {2} (\bibinfo {year} {2016})},\ \Eprint {https://arxiv.org/abs/1511.04455} {arXiv:1511.04455 [astro-ph.GA]} \BibitemShut {NoStop}%
\bibitem [{\citenamefont {Graham}\ \emph {et~al.}(2016)\citenamefont {Graham}, \citenamefont {Durr{\'e}}, \citenamefont {Savorgnan}, \citenamefont {Medling}, \citenamefont {Batcheldor}, \citenamefont {Scott}, \citenamefont {Watson},\ and\ \citenamefont {Marconi}}]{Graham:2016}%
  \BibitemOpen
  \bibfield  {author} {\bibinfo {author} {\bibfnamefont {A.~W.}\ \bibnamefont {Graham}}, \bibinfo {author} {\bibfnamefont {M.}~\bibnamefont {Durr{\'e}}}, \bibinfo {author} {\bibfnamefont {G.~A.~D.}\ \bibnamefont {Savorgnan}}, \bibinfo {author} {\bibfnamefont {A.~M.}\ \bibnamefont {Medling}}, \bibinfo {author} {\bibfnamefont {D.}~\bibnamefont {Batcheldor}}, \bibinfo {author} {\bibfnamefont {N.}~\bibnamefont {Scott}}, \bibinfo {author} {\bibfnamefont {B.}~\bibnamefont {Watson}},\ and\ \bibinfo {author} {\bibfnamefont {A.}~\bibnamefont {Marconi}},\ }\href {https://doi.org/10.3847/0004-637X/819/1/43} {\bibfield  {journal} {\bibinfo  {journal} {Astrophys. J.}\ }\textbf {\bibinfo {volume} {819}},\ \bibinfo {pages} {43} (\bibinfo {year} {2016})},\ \Eprint {https://arxiv.org/abs/1601.05151} {arXiv:1601.05151 [astro-ph.GA]} \BibitemShut {NoStop}%
\bibitem [{\citenamefont {Kormendy}\ and\ \citenamefont {Ho}(2013)}]{Kormendy:2013dxa}%
  \BibitemOpen
  \bibfield  {author} {\bibinfo {author} {\bibfnamefont {J.}~\bibnamefont {Kormendy}}\ and\ \bibinfo {author} {\bibfnamefont {L.~C.}\ \bibnamefont {Ho}},\ }\href {https://doi.org/10.1146/annurev-astro-082708-101811} {\bibfield  {journal} {\bibinfo  {journal} {Ann. Rev. Astron. Astrophys.}\ }\textbf {\bibinfo {volume} {51}},\ \bibinfo {pages} {511} (\bibinfo {year} {2013})},\ \Eprint {https://arxiv.org/abs/1304.7762} {arXiv:1304.7762 [astro-ph.CO]} \BibitemShut {NoStop}%
\bibitem [{\citenamefont {Eby}\ \emph {et~al.}(2022)\citenamefont {Eby}, \citenamefont {Shirai}, \citenamefont {Stadnik},\ and\ \citenamefont {Takhistov}}]{Eby:2021ece}%
  \BibitemOpen
  \bibfield  {author} {\bibinfo {author} {\bibfnamefont {J.}~\bibnamefont {Eby}}, \bibinfo {author} {\bibfnamefont {S.}~\bibnamefont {Shirai}}, \bibinfo {author} {\bibfnamefont {Y.~V.}\ \bibnamefont {Stadnik}},\ and\ \bibinfo {author} {\bibfnamefont {V.}~\bibnamefont {Takhistov}},\ }\href {https://doi.org/10.1016/j.physletb.2021.136858} {\bibfield  {journal} {\bibinfo  {journal} {Phys. Lett. B}\ }\textbf {\bibinfo {volume} {825}},\ \bibinfo {pages} {136858} (\bibinfo {year} {2022})},\ \Eprint {https://arxiv.org/abs/2106.14893} {arXiv:2106.14893 [hep-ph]} \BibitemShut {NoStop}%
\end{thebibliography}%

\end{document}